\documentclass[pra,twocolumn,amsmath,amssymb,floatfix,reprint,footinbib,superscriptaddress,longbibliography,showkeys]{revtex4-1}
\usepackage{dcolumn}
\usepackage{graphicx}
\usepackage{mathrsfs}
\usepackage{mdwlist}
\usepackage{subfigure}
\usepackage{booktabs}
\usepackage{amsmath}
\usepackage{extarrows}
\usepackage{dsfont}
\usepackage{amstext}
\usepackage{amssymb}
\usepackage{amsbsy}
\usepackage{bbm}
\usepackage{bm}
\usepackage{amsthm}
\usepackage{graphicx}
\usepackage{textcomp}
\usepackage{multirow}
\usepackage{color}
\usepackage{diagbox}
\usepackage{slashbox}
\usepackage{sansmath}
\usepackage{makecell}
\usepackage{tabularx}
\usepackage[colorlinks,citecolor=blue]{hyperref}
\definecolor{Dgreen}{RGB}{0, 100, 0}
\usepackage{url}
\usepackage[colorlinks]{hyperref}
\hypersetup{%
	plainpages=true,
	breaklinks=true,  
	hypertexnames=false,  
	pageanchor=true,
	colorlinks=true,
	linkcolor={blue},
	citecolor={blue},
	urlcolor={blue},
	anchorcolor={black}
}

\begin{document}

\title{Two-level systems with periodic $N$-step driving fields: Exact dynamics and quantum state manipulations}

\author{Zhi-Cheng Shi}
\affiliation{Fujian Key Laboratory of Quantum Information and Quantum Optics (Fuzhou University), Fuzhou 350108, China}
\affiliation{Department of Physics, Fuzhou University, Fuzhou 350108, China}

\author{Ye-Hong Chen}
\affiliation{Theoretical Quantum Physics Laboratory, RIKEN Cluster for Pioneering Research, Wako-shi, Saitama 351-0198, Japan}

\author{Wei Qin}
\affiliation{Theoretical Quantum Physics Laboratory, RIKEN Cluster for Pioneering Research, Wako-shi, Saitama 351-0198, Japan}

\author{Yan Xia}\thanks{xia-208@163.com}
\affiliation{Fujian Key Laboratory of Quantum Information and Quantum Optics (Fuzhou University), Fuzhou 350108, China}
\affiliation{Department of Physics, Fuzhou University, Fuzhou 350108, China}

\author{X. X. Yi} \thanks{yixx@nenu.edu.cn}
\affiliation{\mbox{Center for Quantum Sciences and School of Physics, Northeast Normal University, Changchun 130024, China}}

\author{Shi-Biao Zheng}
\affiliation{Fujian Key Laboratory of Quantum Information and Quantum Optics (Fuzhou University), Fuzhou 350108, China}
\affiliation{Department of Physics, Fuzhou University, Fuzhou 350108, China}

\author{Franco Nori}
\affiliation{Theoretical Quantum Physics Laboratory, RIKEN Cluster for Pioneering Research, Wako-shi, Saitama 351-0198, Japan}
\affiliation{Department of Physics, University of Michigan, Ann Arbor, Michigan 48109-1040, USA}

\begin{abstract}

In this work, we derive exact solutions of a dynamical equation, which can represent all two-level Hermitian systems driven by periodic $N$-step driving fields. {For different physical parameters, this dynamical equation displays various phenomena for periodic $N$-step driven systems.}  The time-dependent transition probability can be expressed by a general formula that consists of cosine functions with discrete frequencies, and, remarkably, this formula is suitable for arbitrary parameter regimes.
Moreover, only a few cosine functions (i.e., one to three main frequencies) are sufficient to describe the actual dynamics of the periodic $N$-step driven system.
{Furthermore}, we find that a beating in the transition probability emerges when two (or three) main frequencies are similar.
Some applications are also demonstrated in quantum state manipulations by periodic $N$-step driving fields.

\end{abstract}

\maketitle

\section{Introduction}

Over the past decade, periodically driven systems have become the focus of intensive research, owing to the appearance of several interesting phenomena, including coherent destruction of tunneling \cite{grossmann91,grifoni98}, dynamical freezing and localization \cite{broer04,das10,creffield10,hegde14,nag15,bukov15,agarwala16,luitz18,horstmann07,dai18,choi18}, etc.
Those phenomena are usually inaccessible for undriven systems.
More importantly, periodic driving fields can be exploited to control quantum dynamics, and then perform quantum information processing. For instance, multiphoton resonances can be effectively suppressed by periodic driving fields associated with pulse-shaping techniques \cite{gagnon17a}, and sinusoidal driving fields have been used to prepare entangled states and implement quantum gates \cite{paraoanu06,creffield07,li08,li09,song16,yang19}.  In many-body systems, periodic driving fields also provide a new method for coherent quantum manipulations \cite{eckardt05,eckardt17,sun20}.

As is well known, it is very challenging to obtain the exact dynamics for the general time-dependent Hamiltonian even in the simplest two-level system (TLS) \cite{PhysRevLett.109.060401,PhysRevA.83.033614,Shevchenko12,PhysRevB.86.121303, GonzalezZalba2016,PhysRevB.94.195108,Vandersypen2017,Chatterjee2018,Otxoa2019,Wen2020}.
{Exact solutions} are only acquired in a handful of special cases, such as the well-known Landau-Zener model \cite{landau32,zener32,Shevchenko2010}, Rosen-Zener model \cite{PhysRev.40.502}, Allen-Eberly model \cite{allen1987optical}, etc. The form of the exact solution always contains complicated hypergeometric functions or $\Gamma$ functions.

This similar {difficulty} also exists in time-dependent periodically driven systems. To study the dynamics of periodically driven systems, one usually uses Floquet theory \cite{shirley65,sambe73}.
The basic idea is to deduce the time-independent Floquet (effective) Hamiltonian and the corresponding micromotion operator, and then acquire the effective dynamical evolution of the periodically driven systems.
{Therefore}, some nontrivial physical properties (e.g., dynamical localization \cite{horstmann07,dai18,choi18}) can be qualitatively derived.
Nevertheless, it is worth mentioning that in the process of obtaining the analytical expressions of the Floquet Hamiltonian, one always makes certain approximations for the system parameters, e.g., the high-frequency limit, the weak-coupling regime, etc.
{When these} approximations do not hold, the effective Hamiltonian of periodically driven systems cannot {describe well} the actual dynamics.

Recently, a general formalism \cite{goldman14}, which extends the method introduced in Ref.~\cite{rahav03}, has been proposed to describe periodic-square-wave driven systems \cite{Savelev2002,Cole2006,Savelev2004,Tonomura2006,Shi2016,Ono2019,PhysRevApplied.11.014053}.
More specifically, the dynamics of periodic-square-wave driven systems can be divided into two parts: (i) the long-time-scale dynamics governed by an effective Hamiltonian and (ii) the short-time-scale dynamics governed by a micromotion operator.
Nevertheless, this formalism is only suitable for the off-resonant regime (i.e., the frequency of the driving field is off resonant with any {energy level} separations of the static system) and applicable to a specific square wave that possesses the same time intervals.
Then, this method was generalized to the resonant regime \cite{goldman15} by performing a time-dependent unitary transformation.
However, {this approach} might also be invalid when the frequency of the driving field is arbitrary.

Until now, for arbitrary parameter regimes, exact and analytical solutions of a periodically driven TLS are still difficult to obtain using nonperturbative approaches.
Approximations are often applied in previous works to deal with periodically driven TLSs. For instance, in Ref.~\cite{Gurin2003}, approximations are needed to obtain the analytical expression of the Floquet Hamiltonian.
Here, {our goal is to {obtain} {the accurate effective Hamiltonian and the micromotion operator for an arbitrary parameter regime \emph{without} any approximations}. As a result, we manage to describe the {exact dynamics at arbitrary time} {for periodic $N$-step driven (PNSD) systems}.
{Moreover, we demonstrate that there are different phenomena in PNSD systems when physical parameters satisfy different conditions, as summarized in Table~\ref{fig:0}.}


\renewcommand\arraystretch{1.5}
\begin{table}[t]
	\centering
	\caption{Various phenomena in periodic $N$-step driven systems when physical parameters satisfy different conditions. }
	\label{fig:0}
	\begin{tabular}{lc}
		\hline
		\hline
		 Phenomena ~&~    Conditions \\
        \hline
        \emph{Coherent destruction of tunneling}  &  $\epsilon_{\mathrm{eff}}=0$     \\
        \emph{Complete population transition}  &   $\Delta_{\mathrm{eff}}=0$ \\
        \emph{Periodic evolution}  &   $\cos \mathcal{N}_1\Theta=1$ \\
        \emph{Stepwise evolution}  &   $\cos \mathcal{N}_1\Theta=0$ \\
        \emph{Population swapping}  &  $\cos \Theta=0$  \\
        \emph{Beat phenomenon}  &   $|\omega_{\mathrm{eff}}+\omega_{\mathrm{eff}}^{-}|\gg|\omega_{\mathrm{eff}}-\omega_{\mathrm{eff}}^{-}|$ \\
        \hline
        \hline
	\end{tabular}
\end{table}

In this paper, by deriving the exact expression of the evolution operator, we systematically solve a class of dynamical equations, which can describe {all {two-level systems} driven by periodic $N$-step driving fields}.
Note that the exact solution is valid for the full parameter range since we do not make any approximations.
We {derive} general forms of the amplitudes and phases in the frequency domain. {Moreover,} we find that there always {exist} {a} beat-frequency-like phenomenon in the time-dependent transition probability (hereafter we call it the ``beat phenomenon'') in the PNSD system. In particular, the beat phenomenon can also emerge when adopting resonant pulses with different phases.
We also demonstrate {how to perform phase {measurements}} by using {this} beat phenomenon.

\section{Physical model}

We consider a two-level system interacting with a driving field. The dynamics is governed by the following Hamiltonian ($\hbar=1$)
\begin{eqnarray} \label{1}
H(t)&=&\Upsilon(t)\cdot\bm{\sigma}+\frac{\Delta(t)}{2}\bm{I} \cr\cr
      &=&\Delta(t)|2\rangle\langle 2|+\left[\epsilon(t)e^{i\theta(t)}|1\rangle\langle 2|+\mathrm{H.c.}\right],
\end{eqnarray}
where $$\Upsilon(t)=(\epsilon(t)\cos\theta(t), \epsilon(t)\sin\theta(t),-\Delta(t)/2),$$
$\bm{\sigma}=(\sigma_{x}, \sigma_{y}, \sigma_{z})$ represents the Pauli matrices, and $\bm{I}$ is the identity matrix. The first and second terms of the second line {of Eq.~(\ref{1})} can be regarded as the time-dependent energy bias and tunneling amplitude of the TLS, respectively \cite{grifoni98,Shevchenko2010,Vandersypen2017,Shevchenko12}.

{We do not restrict the values of $\Delta(t)$, $\epsilon(t)$, and $\theta(t)$ here, so this Hamiltonian can describe all Hermitian quantum systems with a two-level structure.} Note that the coefficients $\Delta(t)$, $\epsilon(t)$, and $\theta(t)$ can represent different physical quantities in different systems. For example, $\Delta(t)$ and $\epsilon(t)$ are the quasiparticle momenta in condensed-matter systems \cite{gagnon16,fillion16,gagnon17a}, or the detuning and the Rabi frequency in an atomic system interacting with a laser field \cite{scully97}, or the dc and ac flux biases in superconductor qubits \cite{son09,pan17}, etc. Without loss of generality, we assume that all coefficients are adjustable, and {hereafter we denote} $\Delta(t)$, $\epsilon(t)$, and $\theta(t)$ as the detuning, coupling strength, and phase.

The periodic driving field {studied here} has the form of {a} repeated $N$-step sequence $S_N:\{H_1,H_2,\dots,H_N\}$.
For the $n$th step, the interaction time between the TLS and the driving field is $\tau_n$, and the Hamiltonian $H_{n}$ becomes ($n=1,\dots,N$)
\begin{eqnarray} \label{3}
H_{n}=\Delta_{n}|2\rangle\langle 2|+\left(\epsilon_{n}e^{i\theta_{n}}|1\rangle\langle 2|+\mathrm{H.c.}\right).
\end{eqnarray}
Thus, the period of the $N$-step sequence is
\begin{eqnarray}
T=\sum_{n=1}^{N}\tau_n.
\end{eqnarray}
Note that, {for the sake of clarity}, the time argument
on the parameters are omitted hereafter if these parameters are time independent.

In this work, we also do not restrict the value of each duration $\tau_n$.
As a result, because no quantities in $H(t)$ can be treated as perturbations, both the Floquet theory \cite{shirley65,sambe73} and the rotating-wave approximation \cite{Fuchs2009,silveri17,lambert18,shevchenko18,basak18} are invalid to calculate the analytical expression of the effective Hamiltonian for this class of systems.

\section{Effective Hamiltonian and evolution operator}

Different from the partitioning introduced in Refs.~\cite{rahav03,goldman14}, we directly divide the evolution operator at arbitrary time $t=t'+\mathcal{N}T$ ($t'<T$, $\mathcal{N}=0,1,2,\dots$) into two factors:
\begin{eqnarray}
\mathbb{U}(t)=\mathbb{U}(t')\mathbb{U}(\mathcal{N}T)\equiv \exp[{-i\mathcal{M}(t')}]\exp({-iH_{\mathrm{eff}}\mathcal{N}T}).~~~~
\end{eqnarray}
Physically, the time-dependent micromotion operator $\mathcal{M}(t')$ and the time-independent effective Hamiltonian $H_{\mathrm{eff}}$ describe the short (``fast'' part) and long (``slow'' part) dynamics of the PNSD system, respectively.
The eigenvalues of $H_{\mathrm{eff}}$ are referred as the quasienergies of this system.

After some algebraic {calculations}, we obtain the {expression for the} micromotion operator $\mathcal{M}(t')$ (see Appendix \ref{ia} for details), i.e.,
\begin{equation} \label{2}
\mathcal{M}(t')=\Delta_{\mathcal{M}}(t')|2\rangle\langle 2|+ \left[\epsilon_{\mathcal{M}}(t')e^{i\theta_{\mathcal{M}}(t')}|1\rangle\langle 2|+\mathrm{H.c.}\right],~
\end{equation}
where
\begin{eqnarray}
\Delta_{\mathcal{M}}(t')&=&\frac{2B_n(t')}{\sqrt{1-A_n(t')^2}}\arccos A_n(t'), \nonumber\\[0.8ex]
\epsilon_{\mathcal{M}}(t')&=&\frac{{\sqrt{C_n(t')^2+D_n(t')^2}}}{\sqrt{1-A_n(t')^2}}\arccos A_n(t'), \nonumber\\[0.8ex]
\theta_{\mathcal{M}}(t')&=&\arctan\frac{C_n(t')}{D_n(t')}. \nonumber
\end{eqnarray}
The coefficients $A_n(t')$, $B_n(t')$, $C_n(t')$, and $D_n(t')$ satisfy the following relations:
\begin{eqnarray}
A_{n\!+\!1}\!(t')\!&=&\!A_{n}(\tau'_{n})\!\cos\! E_n(t'\!\!-\!\tau'_{n})\!-\!\left[\mathcal{A}_{n}\!(\tau'_{n})\!\cdot\!\mathcal{E}_n\right]\!{\sin \!E_n(t'\!\!-\!\tau'_{n})},\cr\cr
B_{n\!+\!1}\!(t')\!&=&\!B_{n}(\tau'_{n})\!\cos\! E_n(t'\!\!-\!\tau'_{n})\!+\!\left[\mathcal{B}_{n}\!(\tau'_{n})\!\cdot\! \mathcal{E}_n\right]\!{\sin \!E_n(t'\!\!-\!\tau'_{n})},\cr\cr
C_{n\!+\!1}\!(t')\!&=&\!C_{n}(\tau'_{n})\!\cos\! E_n(t'\!\!-\!\tau'_{n})\!+\!\left[\mathcal{C}_{n}\!(\tau'_{n})\!\cdot\! \mathcal{E}_n\right]\!{\sin \!E_n(t'\!\!-\!\tau'_{n})},\cr\cr
D_{n\!+\!1}\!(t')\!&=&\!D_{n}(\tau'_{n})\!\cos\! E_n(t'\!\!-\!\tau'_{n})\!+\!\left[\mathcal{D}_{n}\!(\tau'_{n})\!\cdot\! \mathcal{E}_n\right]\!{\sin \!E_n(t'\!\!-\!\tau'_{n})},  \nonumber
\end{eqnarray}
where $\tau^{\prime}_{n-1}<t'<\tau^{\prime}_n$, and
\begin{eqnarray}
\tau^{\prime}_n=\sum\nolimits_{k=0}^{n}\tau_k, ~~~~~~\tau^{\prime}_0=0,~~~~~~E_n=\sqrt{\epsilon_n^2+{\Delta_n^2}/{4}}. \nonumber
\end{eqnarray}
The vectors are
\begin{eqnarray}
\mathcal{A}_{n}(t')&=&(D_{n}(t'),C_{n}(t'), B_{n}(t')), \nonumber\\[1ex]
\mathcal{B}_{n}(t')&=&(C_{n}(t'),-D_{n}(t'),A_{n}(t')), \nonumber\\[1ex]
\mathcal{C}_{n}(t')&=&(-B_{n}(t'),A_{n}(t'),D_{n}(t')), \nonumber\\[1ex]
\mathcal{D}_{n}(t')&=&(A_{n}(t'),B_{n}(t'),-C_{n}(t')), \nonumber\\[1ex]
\mathcal{E}_n&=&(\epsilon_n\cos\theta_n, \epsilon_n\sin\theta_n,{\Delta_n}/{2})/{E_n}. \nonumber
\end{eqnarray}

To derive the effective Hamiltonian $H_{\mathrm{eff}}$, we exploit the definition
\begin{eqnarray}
\mathbb{U}(T)\equiv \exp{(-iH_{\mathrm{eff}}T)},
\end{eqnarray}
where $\mathbb{U}(T)$ is the evolution operator within one period. It is found that $H_{\mathrm{eff}}$ is a special case of the micromotion operator $\mathcal{M}(t)$, which satisfies $H_{\mathrm{eff}}=\mathcal{M}(T)/T$. Thus, $H_{\mathrm{eff}}$ is given by
\begin{eqnarray}  \label{5}
H_{\mathrm{eff}}=\Delta_{\mathrm{eff}}|2\rangle\langle 2|+\left(\epsilon_{\mathrm{eff}}e^{i\theta_{\mathrm{eff}}}|1\rangle\langle 2|+\mathrm{H.c.}\right),
\end{eqnarray}
where  $$\{\Delta_{\mathrm{eff}}, \epsilon_{\mathrm{eff}}, \theta_{\mathrm{eff}}\}=\{\Delta_{\mathcal{M}}(T),\epsilon_{\mathcal{M}}(T),\theta_{\mathcal{M}}(T)\}/T.$$
From this effective Hamiltonian, one easily finds that the coherent destruction of tunneling (complete population transition) \cite{grossmann91,grifoni98} would appear in the PNSD system when $\epsilon_{\mathrm{eff}}=0$ ($\Delta_{\mathrm{eff}}=0$); see Appendix \ref{s3} for details.

Working with $\mathbb{U}(t')$ and $\mathbb{U}(T)$, we obtain the exact evolution operator $\mathbb{U}(t)$ for the PNSD system, which is expressed as (see Appendix \ref{ia} for details)
\begin{eqnarray} \label{18a}
\mathbb{U}(t)&=&\left[
                \begin{array}{cc}
                 A(t)+iB(t) & C(t)-iD(t) \\[1ex]
                 -C(t)-iD(t) & A(t)-iB(t) \\
                \end{array}
              \right].
\end{eqnarray}
The coefficients $A(t)$, $B(t)$, $C(t)$, and $D(t)$ are
\begin{widetext}
\begin{eqnarray} \label{s18b}
A(t)&=& A_n(t')\cos\mathcal{N}\Theta-\Big[\mathcal{A}_n(t')\cdot \mathcal{A}^\texttt{T}_N(T)\Big]\frac{\sin\mathcal{N}\!\Theta}{\sin\Theta}, \nonumber\\[1ex]
B(t)&=& B_n(t')\cos\mathcal{N}\Theta+\Big[\mathcal{B}_n(t')\cdot \mathrm{diag}\{-1,-1,1\}\cdot\mathcal{A}^\texttt{T}_N(T)\Big] \frac{\sin\mathcal{N}\Theta}{\sin\Theta}, \nonumber\\[1ex]
C(t)&=& C_n(t')\cos\mathcal{N}\Theta+\Big[\mathcal{C}_n(t')\cdot \mathrm{diag}\{-1,1,-1\}\cdot\mathcal{A}^\texttt{T}_N(T)\Big]\frac{\sin\mathcal{N}\Theta}{\sin\Theta}, \nonumber\\[1ex]
D(t)&=& D_n(t')\cos\mathcal{N}\Theta+\Big[\mathcal{D}_n(t')\cdot \mathrm{diag}\{1,-1,-1\}\cdot\mathcal{A}^\texttt{T}_N(T)\Big] \frac{\sin\mathcal{N}\Theta}{\sin\Theta}, \nonumber
\end{eqnarray}
\end{widetext}
where $\Theta=\arccos A_N(T)$ and the superscript $\texttt{T}$ denotes the transposition.
It is worth mentioning that {we make {no approximations in obtaining the micromotion operator and the effective Hamiltonian}} given by Eqs.~(\ref{2}) and (\ref{5}).
Hence, {the evolution operator} $\mathbb{U}(t)$ given by Eq.~(\ref{18a}) {is \emph{exact and valid} for the full parameter range}.

{Note that {the evolution operator} $\mathbb{U}(t)$ given by Eq.~(\ref{18a}) is {exact} at any time, while it is only valid at some specific {moments} in Ref.~\cite{PhysRevA.52.2245}. This means that the protocol using rectangular pulses in Ref.~\cite{PhysRevA.52.2245} can be regarded as a special case of our results.}
{Additionally, the phenomena of \emph{periodic evolution, stepwise evolution}, and \emph{population swapping} can also be observed in {PNSD systems} when the physical parameters satisfy $\cos \mathcal{N}_1\Theta=1$, $\cos \mathcal{N}_1\Theta=0$, and $\cos \Theta=0$, respectively, where $\mathcal{N}_1$ is an integer; see Appendix \ref{s3} for details.}

In general situations, the first period of the pulse sequence may have a jump, e.g., from $t_0$ to $t_1$ in Fig.~\ref{fig:01}(a).
When this jump is considered, an additional initial kick is needed to describe the dynamics, as discussed in Refs.~\cite{rahav03,goldman14}, {because} this jump might affect the long-time-scale dynamics [see the solid curve in Fig.~\ref{fig:01}(b)].
In contrast, we integrate this jump into the effective Hamiltonian here, and do not require an additional kick. The result is that different starting {times} (phases) of the driving fields would lead to different effective Hamiltonians.

To address this issue more clearly, we employ a quantity $\lambda$ to describe this jump [see Fig.~\ref{fig:01}(a)].
By using another partition of the sequence, we can regard the jump sequence as a complete $(N+1)$-step sequence $S_{N+1}:\{H_m,H_{m+1},\dots,H_N,H_1,\dots,H_m\}$. Here, the final step Hamiltonian is the same to the $m$th step Hamiltonian with the interaction time $(1-\lambda)\tau_1$.
Then, we readily obtain the effective Hamiltonian $H_{\mathrm{eff}}(\lambda)$ for the jump sequences.
It is shown in Fig.~\ref{fig:01}(b) that the long-time-scale dynamics (i.e., the envelope of the solid curve) caused by different starting {times} of the driving fields are well described by the effective Hamiltonian $H_{\mathrm{eff}}(\lambda)$.

\begin{figure} [t]
	\includegraphics[scale=0.49]{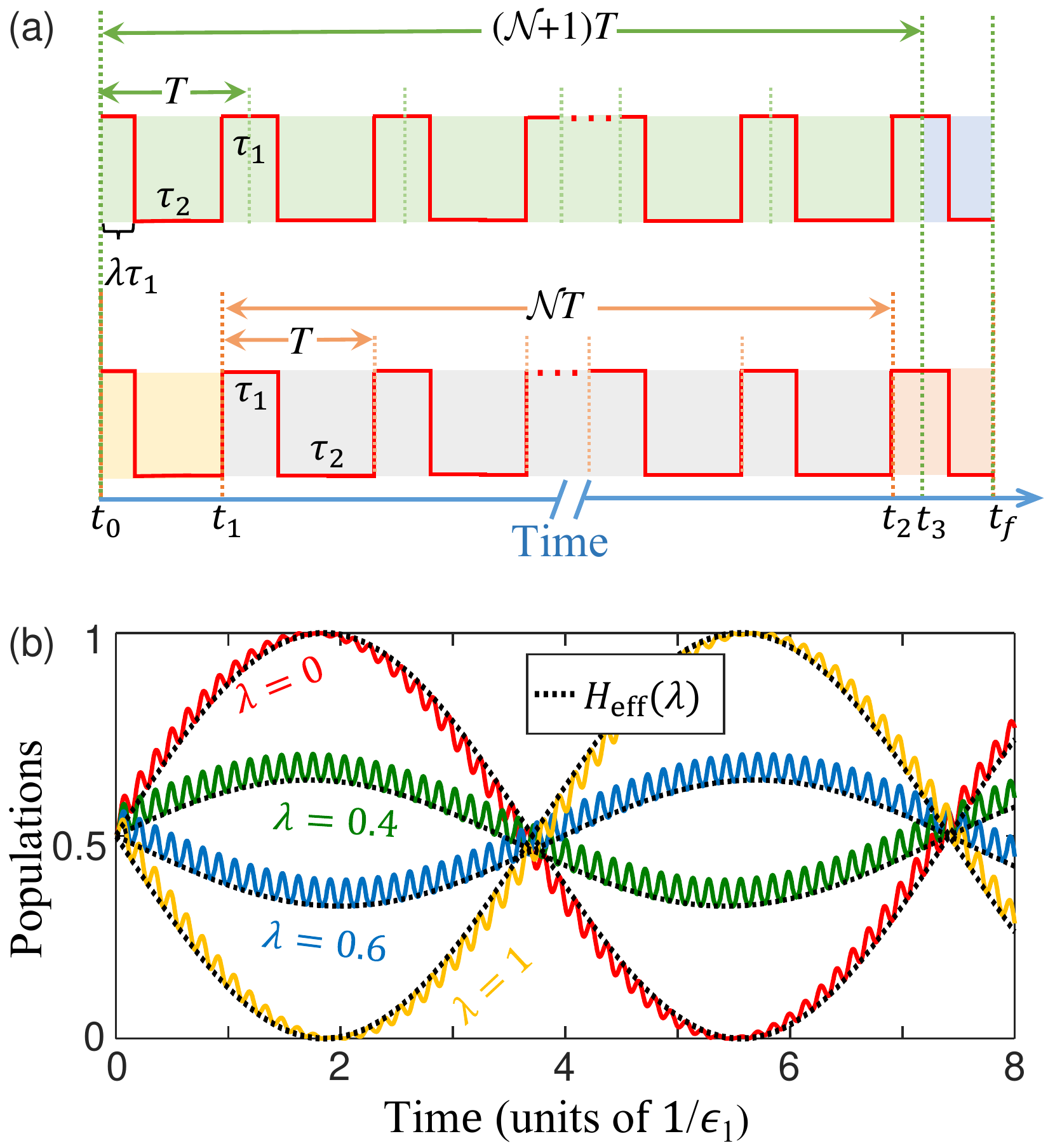}
	\caption{(a) Schematic diagram of distinct partitions of the same two-step sequence with period $T$. The emergence of the first jump is quantified by {the parameter} $\lambda$. In the top panel,
		the evolution operator is divided into two factors (shaded with different colors), $\mathbb{U}(t,t_3)\mathbb{U}(t_3,t_0)$, while it is divided into three factors, $\mathbb{U}(t,t_2)\mathbb{U}(t_2,t_1)\mathbb{U}(t_1,t_0)$ in the bottom panel.
	(b) Population evolution of level $|2\rangle$ for different $\lambda$, where the initial state
	is $(|1\rangle+|2\rangle)/\sqrt{2}$. The black-dotted curves represent the predictions
	given by the effective Hamiltonian $H_{\mathrm{eff}}(\lambda)$. Each dotted curve almost coincides with the envelope of its solid counterpart, indicating that $H_{\mathrm{eff}}(\lambda)$ contains relevant information caused by different starting {times} of the driving fields. The parameters chosen here are $\{\epsilon_{n}/\epsilon_{1}\}=\{1,2\}$,
	$\{\Delta_{n}/\epsilon_{1}\}=\{50,40\}$, and $E_1\tau_1=E_2\tau_2=\pi/2$.}  \label{fig:01}
\end{figure}

\section{Analytical expressions for the transition probability}

Note that there are two fundamental frequencies in this system: the frequency of the $N$-step driving field
\begin{eqnarray}
\omega_T=2\pi/T,
\end{eqnarray}
and the Rabi-like frequency of the effective Hamiltonian
\begin{eqnarray}
\omega_\mathrm{eff}=\sqrt{\epsilon_\mathrm{eff}^2+{\Delta_\mathrm{eff}^2}/{4}}.
\end{eqnarray}
With $\mathbb{U}(t)$ in Eq.~(\ref{18a}), we can express the time-dependent transition probability $P_{12}(t)$ from the state $|1\rangle$ to $|2\rangle$ in terms of cosine functions, yielding
\begin{equation}  \label{7}
P_{12}(t)\!=\!\!\!\sum_{l=-\infty}^{\infty}\!\!\!\Big\{b_{l}\cos\big[(2\omega_{\mathrm{eff}}+l\omega_{T})t-\varphi_l\big] \!+b_{l}^{\prime}\cos\left(l\omega_{T}t\!-\!\varphi_l^{\prime}\right)\!\!\Big\},
\end{equation}
where the amplitudes $\{b_{l},b_{l}^{\prime}\}$ and the phases $\{\varphi_l,\varphi_l^{\prime}\}$ can be {obtained} by the Fourier {transforms} (see Appendix \ref{ie} for details):
\begin{eqnarray}   \label{7a}
b_{l}e^{i\varphi_l}&=&\lim_{K\rightarrow\infty}\frac{2}{KT}\int_0^{KT}\!\!\!\!P_{12}(t)\exp[i(2\omega_{\mathrm{eff}}
+l\omega_{T})t]dt,  \nonumber\\[1ex]
b_{l}^{\prime}e^{i\varphi_l^{\prime}}&=&\lim_{K\rightarrow\infty}\frac{2}{KT}\int_0^{KT}\!\!\!\!P_{12}(t)\exp(il\omega_{T}t)dt.
\end{eqnarray}

From a physical point of view, due to the high-frequency {oscillations}, most amplitudes $\{b_{l},b_{l}^{\prime}\}$ have extremely small values, and thus can be ignored.
{Therefore}, only a few terms in Eq.~(\ref{7}) are kept, but {these are} sufficient to describe the actual dynamics of the PNSD system at arbitrary {times}.

\subsection{Example: Two-step sequence}

\subsubsection{\textbf{Expression for the transition probability}}

Taking the two-step sequence as an example, we can empirically write the expression of the transition probability as
\begin{equation}   \label{8}
P_{12}(t)\!\approx\!P_{12}^{m}(t)\!=\!\frac{1}{2}\Big[ 1\!-\!(1-\lambda)\cos2\omega_{\mathrm{eff}}t\!-\!\lambda\cos2\omega_{\mathrm{eff}}^-t\Big],
\end{equation}
where we set $\Delta_{\mathrm{eff}}=0$, and
\begin{eqnarray}
\omega_{\mathrm{eff}}^{-}=\omega_T/2-\omega_{\mathrm{eff}},~~~~~~~~~~\lambda=p(1-2\vartheta_{1}\vartheta_{2}). \nonumber
\end{eqnarray}
{Here}, $p$ is the maximum transition probability for every Hamiltonian $H_n$, and the {dynamical} phases are $\vartheta_{n}=E_n\tau_n/\pi$.
Note that the focus here is on the dynamical behaviors at {arbitrary} time rather than some {specific} moments (e.g., $t=\mathcal{N}T$) \cite{PhysRevA.52.2245,PhysRevA.55.4418}.
To quantify the validity of Eq.~(\ref{8}) in more detail, we adopt the following definition,
\begin{eqnarray}
\varepsilon^{m}(t_s)=\frac{1}{t_s}\int_{0}^{t_s}\left|P_{12}(t)-P_{12}^{{m}}(t)\right|dt.
\end{eqnarray}
Namely, $\varepsilon^{m}(t_s)$ represents the average error in the time interval $[0,t_s]$ when adopting $P_{12}^{m}(t)$ to describe the actual dynamics.

\begin{figure}[b]
	\centering
	\includegraphics[scale=0.475]{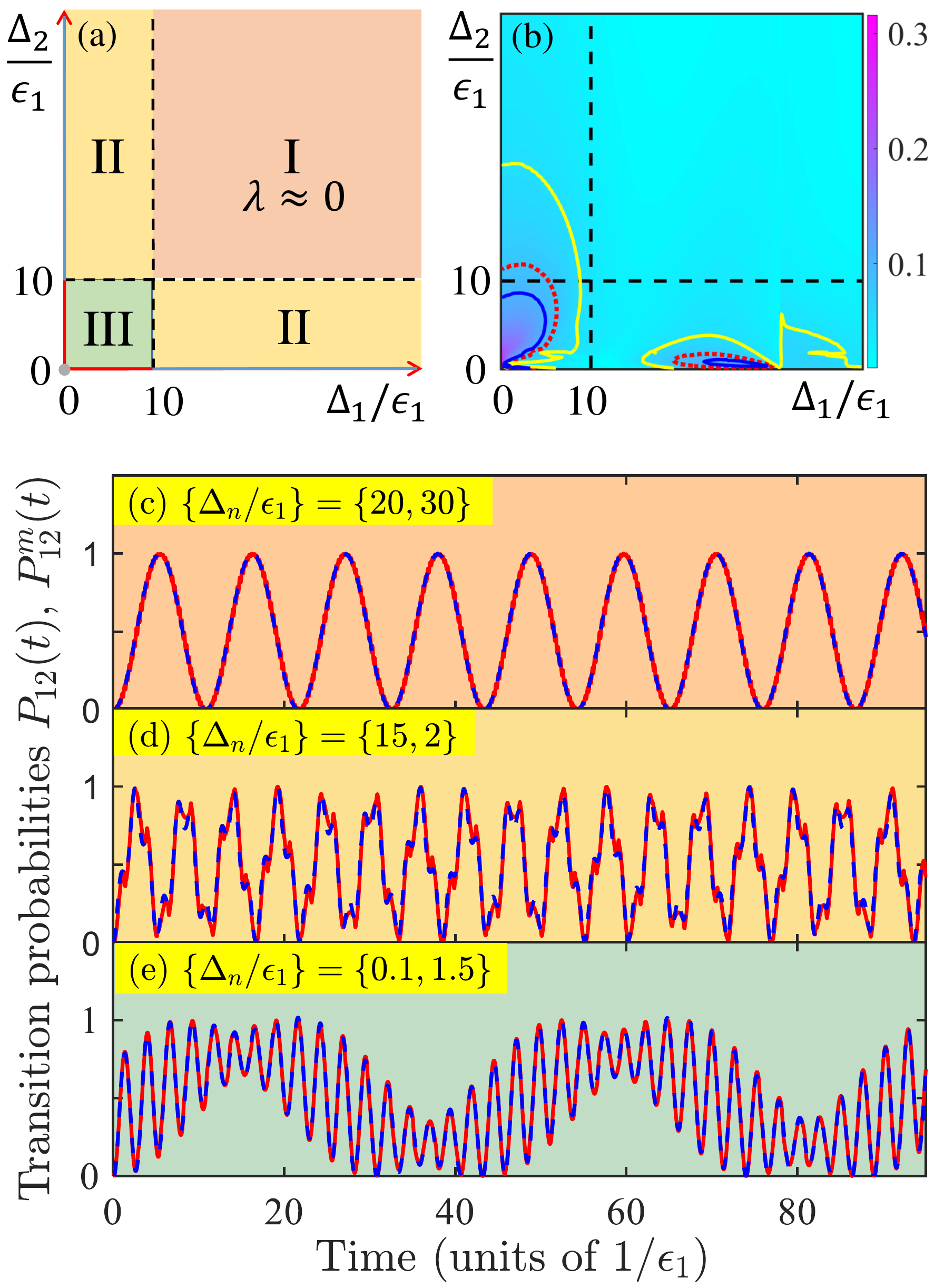}
	\caption{(a) Schematic representation of various parameter regimes. The effective Hamiltonian in Eq.~(\ref{5}) is valid in region I for an arbitrary $\tau_1$. The beat phenomenon can emerge in regions II and III, and is more obvious on the light blue axis or the origin of coordinates ({$\Delta_1=0$ or $\Delta_2=0$}). (b) $\varepsilon^{m}(t_s)$ vs $\Delta_n$, where $\{\theta_n\}=\{0,\pi/3\}$, $\epsilon_2=2\epsilon_1$, $\tau_1=0.2/\epsilon_1$, and $\tau_2$ is chosen to be $\Delta_{\mathrm{eff}}=0$. The yellow solid curves, red dotted curves, and blue solid curves correspond to $\varepsilon^{m}(t_s)=0.05, 0.08, 0.1$, respectively. The results verify the classification in Fig.~\ref{fig:02}(a).
(c, d) Time-dependent transition probabilities $P_{12}(t)$ and $P_{12}^m(t)$ with different detunings in different regions, where the red solid (blue dashed) curves are plotted by $P_{12}(t)$ [$P_{12}^m(t)$]. The appreciable overlap between {these} two curves implies that the actual dynamics at arbitrary time can be well described by $P_{12}^m(t)$. Note that $P_{12}^m(t)$ contains three main frequencies in (e).}  \label{fig:02}
\end{figure}

We plot $\varepsilon^{m}(t_s)$ as a function of the detuning $\Delta_1$ and $\Delta_2$ in Fig.~\ref{fig:02}(b), which confirms the classification in Fig.~\ref{fig:02}(a).
To be more specific, in region I of Fig.~\ref{fig:02}(a), i.e., $\Delta_n/\epsilon_n\gg1$, the actual dynamics of the PNSD system can be well described by the effective Hamiltonian~(\ref{5}) and shows Rabi-like oscillations with the frequency $\omega_{\mathrm{eff}}$, as shown in Fig.~\ref{fig:02}(c).
Note that most {of the previously studied} periodically driven systems \cite{ashhab07,Tuorila10,Neilinger16} are in this regime. But, in contrast to those systems, the validity of $H_{\mathrm{eff}}$ does not depend on the frequency of the driving field in this model \cite{Silveri2015,Ivakhnenko2018}.
In region II of Fig.~\ref{fig:02}(a), the effective Hamiltonian~(\ref{5}) is invalid but the actual dynamics can {still} be well described by Eq.~(\ref{8}), as shown in Fig.~\ref{fig:02}(d).

Note that Eq.~(\ref{8}) might be invalid in some areas of region III, because $\varepsilon^{m}(t_s)$ is sufficiently large.
However, it can be easily solved by including more cosine functions in Eq.~(\ref{8}). For instance, the expression of $P_{12}^m(t)$ keeps three main frequencies (i.e., $\omega_{\mathrm{eff}}$ and $\omega_{\mathrm{eff}}^{\pm}=\omega_T/2\pm\omega_{\mathrm{eff}}$)
in Fig.~\ref{fig:02}(e), where the amplitudes and the phases are achieved by Eq.~(\ref{7a}). More examples can be found in Table \ref{t1} of Appendix \ref{s3}.

\renewcommand\arraystretch{1.5}
\begin{table*}[htbp]
	\centering
	\caption{The physical parameters in Figs.~\ref{sfig:03a}--\ref{sfig:03c} and the corresponding expressions of $P_{12}^m(t)$, where $\{\theta_n\}=\{0,0\}$ and $\epsilon_1=\epsilon_2$.}
	\label{t1}
	\begin{tabular}{cccc}
		\hline
		\hline
		  ~& $\{\epsilon_1\tau_n\}$ ~&~ $\{\Delta_n/\epsilon_1\}$ ~&~    $P_{12}^m(t)$ \\
        \hline
        Fig.~\ref{sfig:03a}(b) &  $\{0.0627,0.1092\}$  &  $\{30,40\}$ &     $0.25-0.25\cos2\omega_{\mathrm{eff}}t$    \\
        
        Fig.~\ref{sfig:03a}(c) &  $\{0.0938,0.1748\}$  &  $\{20,25\}$ &     $0.4-0.4\cos2\omega_{\mathrm{eff}}t$    \\
        
        Fig.~\ref{sfig:03a}(d) &  $\{0.0938,0.1464\}$  &  $\{20,30\}$ &     $0.5-0.5\cos2\omega_{\mathrm{eff}}t$    \\
        
        Fig.~\ref{sfig:03b}(a) &  $\{1.6729,0.2066\}$  &  $\{3,2.19\}$ &     $0.346\!-\!0.1368\cos(2\omega_{\mathrm{eff}}t\!-\!0.3628)\!-\!0.28\cos(2\omega^-_{\mathrm{eff}}t\!-\!0.1416) \!-\!0.06\cos2(\omega_Tt\!-\!0.4846)$    \\
        
        Fig.~\ref{sfig:03b}(b) &  $\{0.3000,1.6461\}$  &  $\{3,1\}$ &     $0.5\!-\!0.3465\cos(2\omega_{\mathrm{eff}}t\!-\!0.1942)\!+\!0.0727\cos(2\omega^-_{\mathrm{eff}}t\!+\!0.54)\!-\!0.2256 \cos2(\omega^+_{\mathrm{eff}}t\!+\!0.1519)$    \\
        
        Fig.~\ref{sfig:03b}(c) &  $\{0.1535,2.1439\}$  &  $\{3,2\}$ &     $0.1972\!+\!0.05357\cos(2\omega_{\mathrm{eff}}t\!+\!0.3389)\!-\!0.1\cos(2\omega_Tt\!-\!0.22)\!-\!0.1524 \cos2(\omega^+_{\mathrm{eff}}t\!+\!0.1185)$    \\
		
        Fig.~\ref{sfig:03c}(a) &  $\{0.3142,3.1377\}$  &  $\{0,40\}$ &     $0.5-0.4919\cos(2\omega_{\mathrm{eff}}t+0.2887)+0.05268\cos(2\omega^-_{\mathrm{eff}}t-0.5747)-0.04523 \cos2\omega^+_{\mathrm{eff}}t$    \\
		
        Fig.~\ref{sfig:03c}(b) &  $\{0.4126,2.5279\}$  &  $\{0,30\}$ &    $0.3375\!-\!0.3241\cos(2\omega_{\mathrm{eff}}t+0.4284)\!-\!0.05774\cos(2\omega^-_{\mathrm{eff}}t-4.016)\!-\!0.04869 \cos2\omega^+_{\mathrm{eff}}t$ \\
		
        Fig.~\ref{sfig:03c}(c) & $\{0.2328,1.5191\}$   & $\{0,50\}$  &     $0.1916\!-\!0.1873\cos(2\omega_{\mathrm{eff}}t+0.3242)\!+\!0.02425\cos(2\omega^-_{\mathrm{eff}}t-0.7375)\!-\!0.02175 \cos2\omega^+_{\mathrm{eff}}t$ \\
        
        Fig.~\ref{sfig:03c}(d) & $\{1.5708,3.1377\}$  &  $\{0,40\}$ &     $0.5\!-\!0.6189\cos(2\omega_{\mathrm{eff}}t+1.048)\!-\!0.06237\cos(2\omega^+_{\mathrm{eff}}t+\omega_Tt-1.044)\!+\!0.1667 \cos2\omega^+_{\mathrm{eff}}t$ \\
        \hline
        \hline
	\end{tabular}
\end{table*}

\subsubsection{\textbf{Beat phenomenon in the transition probability}}

Figure~\ref{fig:02}(b) clearly demonstrates that the actual dynamics of the PNSD {two-level} system can be well described by a superposition of two cosine functions in region II.
An interesting finding is that when the sum of the frequencies of these two cosine functions is much greater than their difference, i.e., $|\omega_{\mathrm{eff}}+\omega_{\mathrm{eff}}^{-}|\gg|\omega_{\mathrm{eff}}-\omega_{\mathrm{eff}}^{-}|$, we can observe
\emph{a beat phenomenon in the transition probability} $P_{12}(t)$, as shown in Figs.~\ref{fig:03a}(a) and \ref{fig:03a}(b).
In particular, {this} beat phenomenon can also emerge for resonant pulses with different phases [the light gray dot in Fig.~\ref{fig:02}(a)], as shown in Figs.~\ref{fig:03a}(c) and \ref{fig:03a}(d).
Furthermore, Figs.~\ref{fig:03a}(e) and \ref{fig:03a}(f) demonstrate that the beat phenomenon emerges in region III of Fig.~\ref{fig:02}(a), where $P_{12}^{m}(t)$ contains three cosine functions (i.e., three similar main frequencies $\omega_{\mathrm{eff}}$ and $\omega_{\mathrm{eff}}^{\pm}$).

\begin{figure}[htbp]
	\centering
	\includegraphics[scale=0.57]{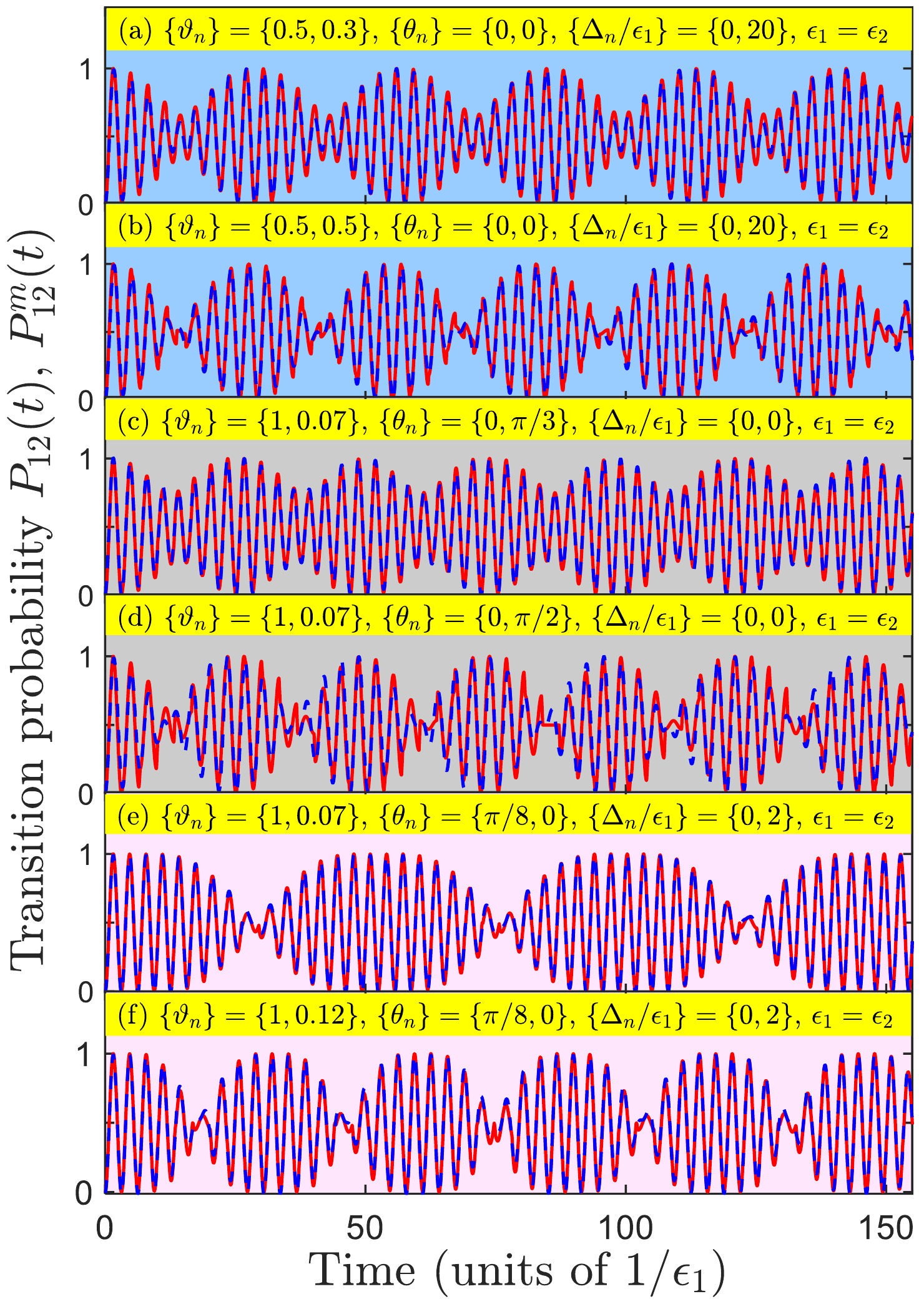}
	\caption{Time-dependent transition probabilities $P_{12}(t)$ and $P_{12}^m(t)$, which show distinct beat phenomena in different regions of Fig.~\ref{fig:02}(a). 
The red solid (blue dashed) curves are plotted by $P_{12}(t)$ [$P_{12}^m(t)$], and the appreciable overlap between these two curves shows the validity of Eq.~(\ref{8}). Note that $P_{12}^m(t)$ contains three main frequencies in (e) and (f).}  \label{fig:03a}
\end{figure}

Physically, {{this} beat phenomenon results from the interference of the phases ($\vartheta_n$ or $\theta_n$), which is different from previous works}, where the quantum beat originates from the interference between different transition channels \cite{Haroche1976,LefebvreBrion2004,Gu2011}.
The beat phenomenon cannot be explained by \emph{only} using the effective Hamiltonian, because the micromotion operator $\mathcal{M}(t')$ plays a key role in the transition process.
Furthermore, the beat phenomenon only emerges in some special physical regimes [e.g., regions II and III in Fig.~\ref{fig:02}(a)], where the traditional methods, such as the Floquet theory \cite{grifoni98,shirley65,sambe73} and the rotating-wave approximation \cite{Fuchs2009,silveri17,lambert18,shevchenko18,basak18}, might not hold.

\subsection{Beat phenomenon in periodic $N$-step driven {systems}}

According to Eq.~(\ref{8}), the beat phenomenon is {more} obvious when $\lambda=1/2$, as shown in Fig.~\ref{fig:03a}(b). Actually, this is the case where the beating always exists in the PNSD system, when at least one Hamiltonian is in the largely detuned (resonant) regime; see Appendix \ref{id} for details.
By modulating the dynamical phases, the transition probability of the PNSD system can be approximately given by
\begin{eqnarray}  \label{9}
P_{12}^{{b}}(t)=\frac{1}{2}\left[\sin^2\varpi_1(t-t_p) +\sin^2\varpi_1'(t-t_p)\right],
\end{eqnarray}
where two similar main frequencies $\varpi_1$ and $\varpi_1'$ are closely related to the effective quantity $\omega_{\mathrm{eff}}$, and $t_p$ is the time-shifting factor.
Moreover, the parity number of $N$ determines the values of the frequencies $\varpi_1$ and $\varpi_1'$.
To be specific, when there are $n_{1}$ Hamiltonians in the resonant regime, we have 
{\fontsize{10pt}{0pt} $$\varpi_1\!=\!\frac{1}{2}\!\left \{
\begin{array}{ll}
    {\!\!\left[n_1\!-\!(-1)^N\right]\omega^{-}_{\mathrm{eff}}}\!+\!\left[n_1\!+\!(-1)^N\right]\omega_{\mathrm{eff}},
    ~~&\textrm{odd} ~n_{1} \nonumber\\[1.5ex]
    {\!\!n_1\omega^{-}_{\mathrm{eff}}+\left[n_1\!-\!2(-1)^N\right]\omega_{\mathrm{eff}}}, & \textrm{even}~n_1 \\
\end{array}
\right. $$
and
 $$\varpi_1'\!=\!\frac{1}{2}\!\left \{
\begin{array}{ll}
    {\!\!(n_1+1)\omega^{-}_{\mathrm{eff}}+\left[n_1\!-\!2(-1)^N\!+\!1\right]\omega_{\mathrm{eff}}}, & \textrm{odd} ~n_{1} \nonumber\\[1.5ex]
    {\!\!\left[n_1\!-\!(\!-\!1)^N\!\!\!+\!1\right]\!\omega^{-}_{\mathrm{eff}} \!+\!\!\left[n_1\!+\!(\!-\!1)^N\!\!\!+\!1\right]\!\omega_{\mathrm{eff}}},  & \textrm{even}~n_1. \\
\end{array}
\right.      $$}

\section{Applications in quantum state manipulations}

\begin{figure}[b]
\centering
\includegraphics[scale=0.53]{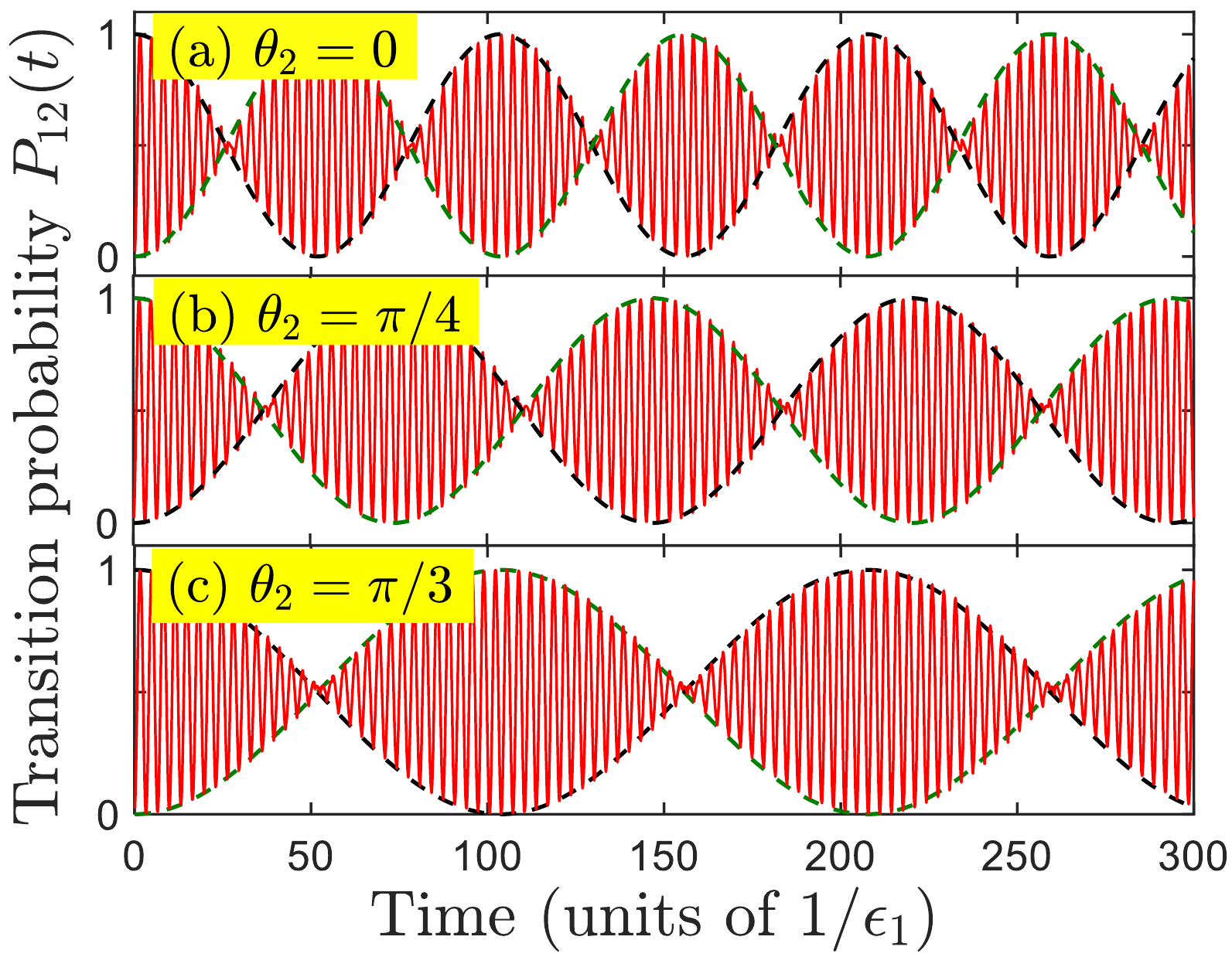}
\caption{Time-dependent transition probability $P_{12}(t)$ in the case of an unknown phase in the two-step driven system, where other parameters are $\theta_1=0$, $\epsilon_1=\epsilon_2$, $\{\Delta_n/\epsilon_1\}=\{0,40\}$, and $\vartheta_n=1/2$. The red solid curves represent the actual dynamics and the {green (black) dashed envelope} curves are {described} by the functions $P_b(t)=1/2(1\pm\cos\omega_bt)$. According to the envelope waveform of $P_{12}(t)$, we can acquire the beat frequency $\omega_b$ by the {green (black) dashed envelope} curves. Different $\omega_b$ reflect  different phases, {given} by Eq.~(\ref{12}).}  \label{fig:3}
\end{figure}

The beat phenomenon observed in the PNSD system could find potential applications in homodyne detection, such as phase {measurements}.
Taking the two-step sequence as an example, we assume that the physical parameters $\{\Delta_n,\epsilon_n\}$ are given and the phase $\theta_n$ is unknown. Then, each duration can be chosen to be $\tau_n=\pi/2E_n$. In this situation, the beating in the transition probability would appear in this system, as shown in Fig.~\ref{fig:3}.
According to Eq.~(\ref{9}), the physical parameters are closely interconnected, and have the relations ($N=2$) 
\begin{eqnarray}  \label{12}
\omega_b=\varpi_1'-\varpi_1\approx\frac{2\epsilon_2}{T E_2}\cos(\theta_1-\theta_2).
\end{eqnarray}
Thus, we can map the phase $\theta_n$ onto the beat frequency, enabling it to be easily measured by detecting the envelope waveform of the beat signal.

In Fig.~\ref{fig:3}, we illustrate the time-dependent transition probability in the case of an unknown phase. {These} results demonstrate that the dynamical behaviors are different and we can acquire the beat frequency $\omega_b$ by population measurements. Then, by {inversely} solving Eq.~(\ref{12}), one can obtain the phase of the system.

\begin{figure}
	\centering
	\includegraphics[scale=0.37]{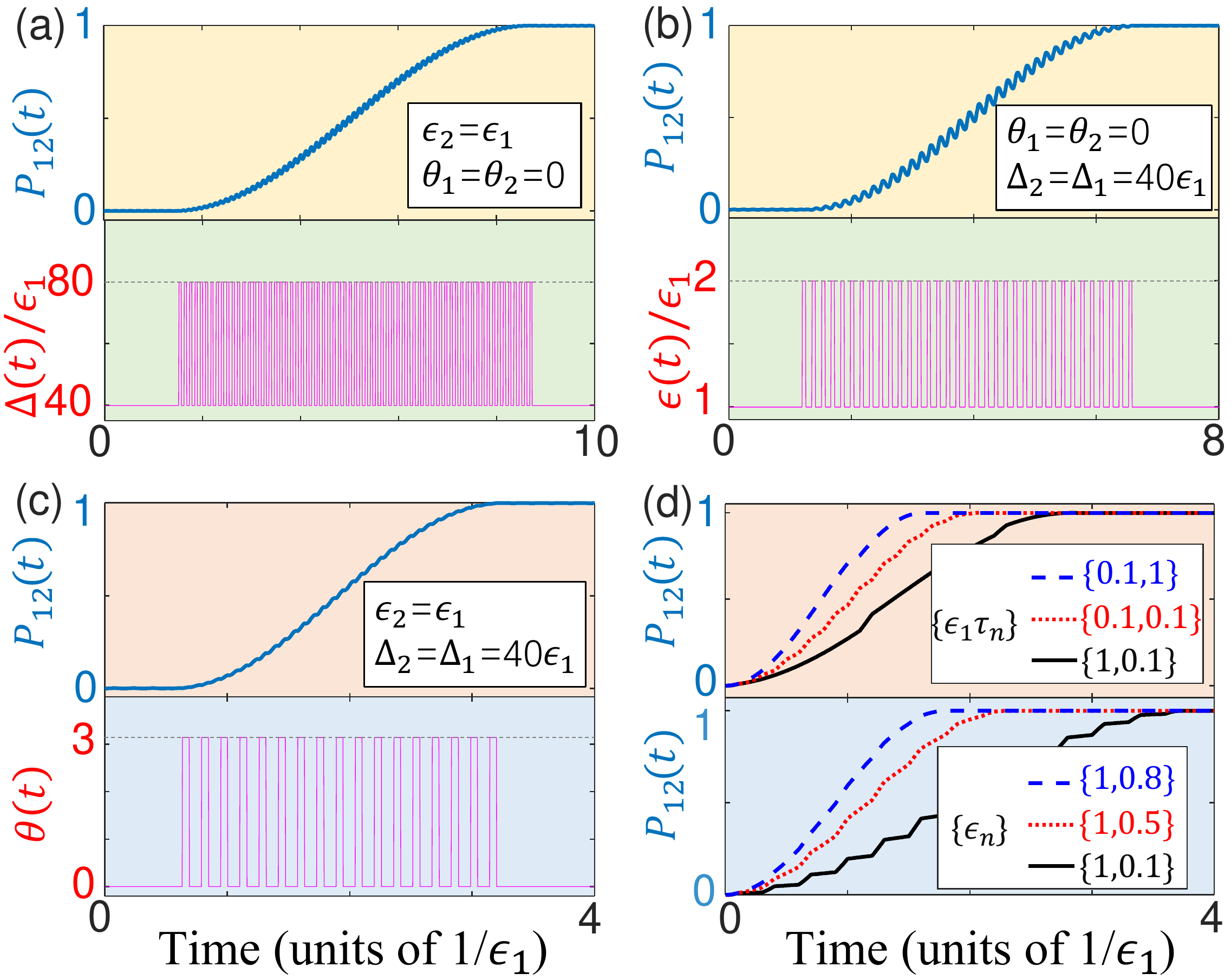}
	\caption{ Time-dependent transition probability $P_{12}(t)$ and the corresponding
	waveform of physical quantities: (a) detuning modulation, (b) coupling strength
		modulation, and (c) phase modulation. In (a), (b), and (c), by merely modulating one parameter with a square-wave sequence, we achieve  the population inversion in the TLS, respectively. (d) Population inversion by merely modulating $\{\tau_{n}\}$ (upper panel, $\epsilon_2=\epsilon_{1}/2$) or $\epsilon_{2}$ (lower panel, $\{\tau_{n}\}=\{0.1,0.2\}/\epsilon_{1}$). Panel (d) shows that the transition time is adjustable by choosing different quantities in the resonant pulses.}  \label{fig:05}
\end{figure}

The periodic $N$-step driving field can be widely used in quantum coherent control, since all physical quantities can be exploited to manipulate quantum states.
For example, to implement {the} \emph{complete transition} from $|1\rangle$ to $|2\rangle$, we can adopt multiple choices to design physical parameters in the PNSD system, such as the detuning $\Delta(t)$ [see the bottom panel of Fig.~\ref{fig:05}(a)], the coupling strength $\epsilon(t)$ [see the bottom panel of Fig.~\ref{fig:05}(b)], or the phase $\theta(t)$ [see the bottom panel of Fig.~\ref{fig:05}(c)].

This method is particularly useful for complicated quantum systems.
Furthermore, for resonant pulses with different coupling strengths, the total evolution time of {the} complete transition reads
\begin{eqnarray}
\mathbb{T}=\frac{\pi(\tau_1+\tau_2)}{2(\epsilon_1\tau_1+\epsilon_2\tau_2)}, ~~~\tau_n<\frac{\pi}{2\epsilon_n}, ~~~n=1,2.
\end{eqnarray}
Here, ${\pi}/{2\epsilon_1}\leq\mathbb{T}\leq {\pi}/{2\epsilon_2}$, when $\epsilon_2<\epsilon_1$.
By modulating the interaction time $\tau_n$ or the coupling strength $\epsilon_n$, we can control the transition time from $|1\rangle$ to $|2\rangle$, as shown in Fig.~\ref{fig:05}(d).

Finally, the \emph{coherent destruction of tunneling} phenomenon in PNSD systems offers us a possible way to implement the forbidden transition by resonant pulses. The physical parameters should satisfy $\theta_2=\theta_1+\pi$ and $\epsilon_2\tau_2=\epsilon_1\tau_1\ll{\pi}/{2}$.

\section{Conclusion}

We have presented the exact solution of the evolution operator for the full parameter range in the periodic $N$-step driven TLS. {Then, we display different physical parameter regimes for various phenomena in this system,
including the well-known coherent destruction of tunneling, complete population transition, and beat phenomenon.}

The time-dependent transition probability of the PNSD system can be expanded by a few cosine functions with discrete frequencies. {Generally speaking, {no} more than three main frequencies are {needed} to describe the actual dynamics of the PNSD system.}
{In addition, we derive the exact expression of the effective Hamiltonian of the PNSD system, which is valid in the largely detuned regime (i.e., $\Delta_n/\epsilon_n\gg1$) or the high-frequency limit.
When the micromotion operator contributes to the evolution, the effective Hamiltonian would be invalid. For this situation, we introduce a modified transition probability to describe the actual dynamics.}

Moreover, we have demonstrated that the beat phenomenon always exists in the $N$-step driven TLS, which cannot be explained by only the effective Hamiltonian.
We have also {described} the beat phenomenon by the resonant pulses with different phases in the PNSD system, and given the general expressions of two similar frequencies when $N>2$.
The beat phenomenon originates from the interference of the phases and {can be} helpful for a driven quantum TLS used in quantum state manipulations.
\\

\begin{acknowledgments}
	We acknowledge helpful discussions with Dr.~S.~N.~Shevchenko and Dr.~Y.~K.~Jiang.
	This work is supported by the National Natural Science Foundation of China under Grant No. 11805036, No. 11674060, No. 11534002, and No. 11775048, the Natural Science Funds for Distinguished Young Scholar of Fujian Province under Grant No. 2020J06011, and the Project from Fuzhou University under Grant No. JG202001-2.
	Y.-H.C. is supported by the Japan Society for the Promotion of Science (JSPS) KAKENHI Grant No.~JP19F19028.
	 F.N. is supported in part by:
	Nippon Telegraph and Telephone Corporation (NTT) Research,
	the Japan Science and Technology Agency (JST) [via
	the Quantum Leap Flagship Program (Q-LEAP),
	the Moonshot R\&D Grant No.~JPMJMS2061, and
	the Centers of Research Excellence in Science and Technology (CREST) Grant No.~JPMJCR1676],
	the Japan Society for the Promotion of Science (JSPS)
	[via the Grants-in-Aid for Scientific Research (KAKENHI) Grant No.~JP20H00134 and the
	JSPS–RFBR Grant No. JPJSBP120194828],
	the Army Research Office (ARO) (Grant No.~W911NF-18-1-0358),
	the Asian Office of Aerospace Research and Development (AOARD) (via Grant No.~FA2386-20-1-4069), and
	the Foundational Questions Institute Fund (FQXi) via Grant No.~FQXi-IAF19-06.
\end{acknowledgments}

\begin{appendix}
\begin{widetext}

\section{Solutions for the dynamical equation in periodic $N$-step driven two-level systems}  \label{ia}

\subsection{General formalism}

The physical model we consider here is a two-level system (TLS) driven by a periodic $N$-step driving field. The general form of its Hamiltonian reads ($\hbar=1$)
\begin{eqnarray} \label{s1}
H(t)&=&\sum_{k=0}^{\mathcal{N}}\sum_{n=1}^{N}H_n\left[\zeta\left(t-kT-\sum_{m=0}^{n-1}\tau_m\right)-\zeta\left(t-kT -\sum_{m=0}^{n}\tau_m\right)\right], \nonumber\\[1.2ex]
H_{n}&=&\Delta_{n}|2\rangle\langle 2|+\epsilon_{n}\exp({i\theta_{n}})|1\rangle\langle 2|+\mathrm{H.c.},
\end{eqnarray}
where $\zeta(t)$ is the Heaviside function, $\tau_0=0$, $\tau_n$ ($n=1,\dots,N$) is the interaction time between the TLS and the $n$th-step driving field, and $T=\tau_N+\tau_{N-1}+\cdots+\tau_1$ is the period of the $N$-step sequence.

Generally speaking, the evolution operator of periodically driven systems can be divided into two parts:
\begin{eqnarray}
\mathbb{U}(t)=\mathbb{P}(t)\exp({-iH_{\mathrm{eff}}t}),
\end{eqnarray}
where the unitary operator $\mathbb{P}(t)$ has the same period as the system Hamiltonian.
The eigenvalues of the effective Hamiltonian $H_{\mathrm{eff}}$ are referred to as the quasienergies of the system.
Different from both this partitioning and the partition introduced in Refs. \cite{rahav03,goldman14}, we directly divide the evolution operator of a periodic $N$-step driven system at arbitrary final time $t=t'+\mathcal{N}T$ ($t'<T$, $\mathcal{N}=1,2,\dots$) into two parts:
\begin{eqnarray}
\mathbb{U}(t)=\mathbb{U}(t')\underbrace{\mathbb{U}(T)\cdots\mathbb{U}(T)}_{\mathcal{N}}= \mathbb{U}(t')\mathbb{U}(\mathcal{N}T)\equiv \exp[{-i\mathcal{M}(t')}]\exp({-iH_{\mathrm{eff}}\mathcal{N}T}),
\end{eqnarray}
where $$\mathbb{U}(T)=U_N(\tau_N)U_{N-1}(\tau_{N-1})\cdots U_1(\tau_1)\equiv \exp({-iH_{\mathrm{eff}}T})$$ is the evolution operator within one period. $U_n(t')=\exp({-iH_nt'})$ represents the evolution operator for the $n$th Hamiltonian, $n=1,\dots,N$.
{Physically, the time-dependent micromotion operator $\mathcal{M}(t')$ describes the short-time-scale (``fast'' part) dynamics of the system, and the time-independent effective Hamiltonian $H_{\mathrm{eff}}$ describes the long-time-scale (``slow'' part) dynamics of the system}.

In order to derive the expression of the evolution operator $\mathbb{U}(t)$ for the periodic $N$-step driven system, we need to calculate both the micromotion operator $\mathcal{M}(t')$ and the effective Hamiltonian $H_{\mathrm{eff}}$.
To be specific, the micromotion operator $\mathcal{M}(t')=\Delta_{\mathcal{M}}(t')|2\rangle\langle 2|+\epsilon_{\mathcal{M}}(t')\exp[{i\theta_{\mathcal{M}}(t')}]|1\rangle\langle 2|+\mathrm{H.c.}$ is defined by the evolution operator $\mathbb{U}(t')$, such that
\begin{eqnarray}
\exp[{-i\mathcal{M}(t')}]\equiv\mathbb{U}(t')=\left \{
\begin{array}{lllll}
    U_1(t'),  &t'\in[0,\tau_1] \\[0.7ex]
    U_2(t'-\tau_1)U_1(\tau_1), &t'\in(\tau_1,\tau_2+\tau_1] \\[0.7ex]
    U_3(t'-\tau_2-\tau_1)U_2(\tau_2)U_1(\tau_1),  &t'\in(\tau_2+\tau_1,\tau_3+\tau_2+\tau_1] \\[0.7ex]
    \cdots &\cdots  \\[0.7ex]
    U_N(t'-\sum\nolimits_{k=1}^{N-1}\tau_k)U_{N-1}(\tau_{N-1})\cdots U_2(\tau_2)U_1(\tau_1), ~~~~~~&t'\in(\sum\nolimits_{k=1}^{N-1}\tau_k,\sum\nolimits_{k=1}^{N}\tau_k],~~~ \\[0.7ex]
\end{array}
\right.
\end{eqnarray}
where $U_n(t')=\exp{(-iH_nt')}$.

When $t'\in[0,\tau_1]$, the evolution operator becomes
\begin{eqnarray}  \label{s16}
\mathbb{U}(t')\!=\!
U_1(t')\!=\!\!\left[\!\!
                \begin{array}{cc}
                 \displaystyle\cos(E_1t')+i\frac{\Delta_1}{2E_1}\sin(E_1t') & \displaystyle-i\frac{\epsilon_1e^{i\theta_1}}{E_1}\sin(E_1t') \\[2ex]
                  \displaystyle-i\frac{\epsilon_1e^{-i\theta_1}}{E_1}\sin(E_1t') & \displaystyle\cos(E_1t')-i\frac{\Delta_1}{2E_1}\sin(E_1t') \\
                \end{array}
              \!\!\right]\!\!=\!\!\left[\!\!
                \begin{array}{cc}
                 A_1(t')+iB_1(t') & C_1(t')-iD_1(t') \\[1ex]
                 -C_1(t')-iD_1(t') & A_1(t')-iB_1(t') \\
                \end{array}
              \!\!\right]\!\!.~~~~~~
\end{eqnarray}
Comparing the left-hand side and the right-hand side of Eq.~(\ref{s16}), we find
$$A_1(t')=\cos(E_1t'),~~~~~\ B_1(t')=\frac{\Delta_1}{2E_1}\sin(E_1t'),~~~~~\ C_1(t')=\frac{\epsilon_1\sin\theta_1}{E_1}\sin(E_1t'),
~~~~~\ D_1(t')=\frac{\epsilon_1\cos\theta_1}{E_1}\sin(E_1t'),$$ where $E_1=\sqrt{\epsilon_1^2+{\Delta_1^2}/{4}}$.
Then, by inversely solving the equation $\exp{[-i\mathcal{M}(t')]}=\mathbb{U}(t')$, we have
$$\Delta_{\mathcal{M}}(t')=\frac{2B_1(t')}{\sqrt{1-A_1(t')^2}}\arccos A_1(t'), ~~~~~\ \epsilon_{\mathcal{M}}(t')=\frac{\sqrt{C_1(t')^2+D_1(t')^2}}{\sqrt{1-A_1(t')^2}}\arccos A_1(t'),~~~~~ \ \theta_{\mathcal{M}}(t')=\arctan\frac{C_1(t')}{D_1(t')}.$$

Similarly, when $t'\in(\tau_1,\tau_2+\tau_1]$, the evolution operator is
\begin{eqnarray}  \label{s17}
\mathbb{U}(t')&=&U_2(t'-\tau_1)U_1(\tau_1)=\exp[{-iH_2(t'-\tau_1)}]\exp({-iH_1\tau_1})  \cr\cr
             &=&\!\!\left[\!
                \begin{array}{cc}
                \displaystyle \cos E_2(t'\!\!-\!\tau_1)+i\frac{\Delta_2}{2E_2}\sin E_2(t'\!\!-\!\tau_1) & \displaystyle-i\frac{\epsilon_2e^{i\theta_2}}{E_2}\sin E_2(t'-\tau_1) \\[1ex]
                 \displaystyle-i\frac{\epsilon_2e^{-i\theta_2}}{E_2}\sin E_2(t'-\tau_1) &
                \!\!\!\!\!\!  \displaystyle\cos E_2(t'\!\!-\!\tau_1)-i\frac{\Delta_2}{2E_2}\sin E_2(t'\!\!-\!\tau_1) \\
                \end{array}
              \!\!\!\right]\!\!\cdot\!\!\left[\!\!
                \begin{array}{cc}
                 A_1(\tau_1)+iB_1(\tau_1) & C_1(\tau_1)-iD_1(\tau_1) \\[1ex]
                 -C_1(\tau_1)-iD_1(\tau_1) & A_1(\tau_1)-iB_1(\tau_1) \\
                \end{array}
              \!\!\right]  \cr\cr\cr
             &=&\left[
                \begin{array}{cc}
                 A_2(t')+iB_2(t') & C_2(t')-iD_2(t') \\[1ex]
                 -C_2(t')-iD_2(t') & A_2(t')-iB_2(t') \\
                \end{array}
              \right].
\end{eqnarray}
According to Eq.~(\ref{s17}), we obtain the coefficients $A_2(t')$, $B_2(t')$, $C_2(t')$, and $D_2(t')$
\begin{eqnarray}
A_2(t')&=&\cos E_1\tau_1\cos E_2(t'-\tau_1)-\frac{\Delta_1\Delta_2+4\epsilon_1\epsilon_2\cos(\theta_1-\theta_2)}{4E_1E_2}\sin E_1\tau_1\sin E_2(t'-\tau_1), \cr\cr
B_2(t')&=&\frac{\Delta_1}{2E_1}\sin E_1\tau_1\cos E_2(t'-\tau_1)+\left[\frac{\Delta_2}{2E_2}\cos E_1\tau_1 +\frac{\epsilon_1\epsilon_2\sin(\theta_1-\theta_2)}{E_1E_2}\sin E_1\tau_1\right]\sin E_2(t'-\tau_1),      \cr\cr
C_2(t')&=&\frac{\epsilon_1\sin\theta_1}{E_1}\sin E_1\tau_1\cos E_2(t'-\tau_1)+\left(\frac{\epsilon_2\sin\theta_2}{E_2}\cos E_1\tau_1+\frac{\Delta_1\epsilon_2\cos\theta_2-\Delta_2\epsilon_1\cos\theta_1}{2E_1E_2}\sin E_1\tau_1\right)\sin E_2(t'-\tau_1),      \cr\cr
D_2(t')&=&\frac{\epsilon_1\cos\theta_1}{E_1}\sin E_1\tau_1\cos E_2(t'-\tau_1)+\left(\frac{\epsilon_2\cos\theta_2}{E_2}\cos E_1\tau_1+\frac{\Delta_1\epsilon_2\sin\theta_2-\Delta_2\epsilon_1\sin\theta_1}{2E_1E_2}\sin E_1\tau_1\right)\sin E_2(t'-\tau_1). \nonumber
\end{eqnarray}
As a result, we have
$$\Delta_{\mathcal{M}}(t')=\frac{2B_2(t')}{\sqrt{1-A_2(t')^2}}\arccos A_2(t'),~~~~~\ \epsilon_{\mathcal{M}}(t')=\frac{\sqrt{C_2(t')^2+D_2(t')^2}}{\sqrt{1-A_2(t')^2}}\arccos A_2(t'),~~~~~\ \theta_{\mathcal{M}}(t')=\arctan\frac{C_2(t')}{D_2(t')}.$$

It is easily found from Eqs.~(\ref{s16}) and (\ref{s17}) that the matrix $\mathbb{U}(t')$ has definite symmetry: the real (imaginary) parts of the diagonal (off-diagonal) elements are identical, while the imaginary (real) parts of the diagonal (off-diagonal) elements are opposite.
Thus, we assume further that the evolution operator $\mathbb{U}(\tau'_{n-1})$ after the $(n-1)$-step sequence can be written as
\begin{eqnarray}
\mathbb{U}(\tau'_{n-1})=U_{n-1}(\tau_{n-1})\cdots U_2(\tau_2)U_1(\tau_1)
              =\left[
                \begin{array}{cc}
                 A_{n-1}(\tau'_{n-1})+iB_{n-1}(\tau'_{n-1}) & C_{n-1}(\tau'_{n-1})-iD_{n-1}(\tau'_{n-1}) \\[1ex]
                 -C_{n-1}(\tau'_{n-1})-iD_{n-1}(\tau'_{n-1}) & A_{n-1}(\tau'_{n-1})-iB_{n-1}(\tau'_{n-1}) \\
                \end{array}
              \right],
\end{eqnarray}
where $\tau'_{n}=\sum\nolimits_{k=1}^{n}\tau_k$.
When $\tau'_{n-1}<t'<\tau'_{n}$, the evolution operator $\mathbb{U}(t')$ is
\begin{eqnarray}   \label{s18}
\mathbb{U}(t')&=&U_n(t'-\tau'_{n-1})\mathbb{U}(\tau'_{n-1}) \cr\cr
             &=&\left[
                \begin{array}{cc}
                \displaystyle\cos E_n(t'-\tau'_{n-1})+i\frac{\Delta_n}{2E_n}\sin E_n(t'-\tau'_{n-1}) & \displaystyle -i\frac{\epsilon_ne^{i\theta_n}}{E_n}\sin E_n(t'-\tau'_{n-1}) \\[1ex]
                \displaystyle -i\frac{\epsilon_ne^{-i\theta_n}}{E_n}\sin E_n(t'-\tau'_{n-1}) &  \cos \displaystyle E_n(t'-\tau'_{n-1})-i\frac{\Delta_n}{2E_n}\sin E_n(t'-\tau'_{n-1}) \\
                \end{array}
              \right]\cr\cr\cr
              &&\cdot\left[
                \begin{array}{cc}
                 A_{n-1}(\tau'_{n-1})+iB_{n-1}(\tau'_{n-1}) & C_{n-1}(\tau'_{n-1})-iD_{n-1}(\tau'_{n-1}) \\[1ex]
                 -C_{n-1}(\tau'_{n-1})-iD_{n-1}(\tau'_{n-1}) & A_{n-1}(\tau'_{n-1})-iB_{n-1}(\tau'_{n-1}) \\
                \end{array}
              \right]     \cr\cr\cr
              &=&\left[
                \begin{array}{cc}
                 A_n(t')+iB_n(t') & C_n(t')-iD_n(t') \\[1ex]
                 -C_n(t')-iD_n(t') & A_n(t')-iB_n(t') \\
                \end{array}
              \right].
\end{eqnarray}
According to Eq.~(\ref{s18}), we obtain the following recursive equations for the coefficients $A_n(t')$, $B_n(t')$, $C_n(t')$, and $D_n(t')$:
\begin{eqnarray}  \label{s19}
A_n(t')&=&A_{n-1}(\tau'_{n-1})\cos E_n(t'-\tau'_{n-1})-\left[\mathcal{A}_{n-1}(\tau'_{n-1})\cdot\mathcal{E}_n\right]{\sin E_n(t'-\tau'_{n-1})},\cr\cr
B_n(t')&=&B_{n-1}(\tau'_{n-1})\cos E_n(t'-\tau'_{n-1})+\left[\mathcal{B}_{n-1}(\tau'_{n-1})\cdot \mathcal{E}_n\right]{\sin E_n(t'-\tau'_{n-1})},\cr\cr
C_n(t')&=&C_{n-1}(\tau'_{n-1})\cos E_n(t'-\tau'_{n-1})+\left[\mathcal{C}_{n-1}(\tau'_{n-1})\cdot \mathcal{E}_n\right]{\sin E_n(t'-\tau'_{n-1})},\cr\cr
D_n(t')&=&D_{n-1}(\tau'_{n-1})\cos E_n(t'-\tau'_{n-1})+\left[\mathcal{D}_{n-1}(\tau'_{n-1})\cdot \mathcal{E}_n\right]{\sin E_n(t'-\tau'_{n-1})},
\end{eqnarray}
where the vectors are
\begin{eqnarray}
  \mathcal{A}_{n-1}(t')&=&(D_{n-1}(t'),C_{n-1}(t'), B_{n-1}(t')),\cr\cr
  \mathcal{B}_{n-1}(t')&=&(C_{n-1}(t'),-D_{n-1}(t'),A_{n-1}(t')),\cr\cr \mathcal{C}_{n-1}(t')&=&(-B_{n-1}(t'),A_{n-1}(t'),D_{n-1}(t')),\cr\cr
  \mathcal{D}_{n-1}(t')&=&(A_{n-1}(t'),B_{n-1}(t'),-C_{n-1}(t')),\cr\cr
  \mathcal{E}_n&=&(\epsilon_n\cos\theta_n, \epsilon_n\sin\theta_n,{\Delta_n}/{2})/{E_n},  \nonumber
\end{eqnarray}
and $E_n=\sqrt{\epsilon_n^2+{\Delta_n^2}/{4}}$.
As a result, we have
\begin{eqnarray} \label{s18c}
\Delta_{\mathcal{M}}(t')=\frac{2B_n(t')}{\sqrt{1-A_n(t')^2}}\arccos A_n(t'),~~\ \epsilon_{\mathcal{M}}(t')=\frac{\sqrt{C_n(t')^2+D_n(t')^2}}{\sqrt{1-A_n(t')^2}}\arccos A_n(t'), ~~\ \theta_{\mathcal{M}}(t')=\arctan\frac{C_n(t')}{D_n(t')}.~~~~~~~
\end{eqnarray}
These are the exact solutions for the micromotion operator $\mathcal{M}(t')$ when $\tau'_{n-1}<t'<\tau'_{n}$, $n=1,\dots,N-1$.

To derive the effective Hamiltonian, we need to {use} the definition
$$\mathbb{U}(T)=U_N(\tau_N)U_{N-1}(\tau_{N-1})\cdots U_2(\tau_2)U_1(\tau_1)\equiv \exp{(-iH_{\mathrm{eff}}T)}.$$
We further find that the effective Hamiltonian $H_{\mathrm{eff}}$ is a special case of the micromotion operator $\mathcal{M}(t)$, and it satisfies the relation: $H_{\mathrm{eff}}={\mathcal{M}(T)}/{T}$. Therefore, the general form of the time-independent effective Hamiltonian $H_{\mathrm{eff}}$ for the periodic $N$-step driven system reads
\begin{eqnarray}  \label{s5}
H_{\mathrm{eff}}=\Delta_{\mathrm{eff}}|2\rangle\langle 2|+\big[\epsilon_{\mathrm{eff}}\exp({i\theta_{\mathrm{eff}}})|1\rangle\langle 2|+\mathrm{H.c.}\big],
\end{eqnarray}
where the effective quantities are
$$\Delta_{\mathrm{eff}}=\frac{2B_N(T)}{T\sqrt{1-A_N(T)^2}}\arccos A_N(T),~~~~~\ \epsilon_{\mathrm{eff}}=\frac{\sqrt{C_N(T)^2+D_N(T)^2}}{T\sqrt{1-A_N(T)^2}}\arccos A_N(T),
~~~~~\ \theta_{\mathrm{eff}}=\arctan\frac{C_N(T)}{D_N(T)}.$$
Note that the coefficient $A_N(T)$ should be positive when the transition probability $P_{12}(t)$ from the state $|1\rangle$ to $|2\rangle$ during the time interval $[0,T]$ is less than $P_{12}\Big(\displaystyle\frac{\pi}{2\omega_{\mathrm{eff}}}\Big)$, where $\omega_\mathrm{eff}=\sqrt{\epsilon_\mathrm{eff}^2+{\Delta_\mathrm{eff}^2}/{4}}$.

Now, we can calculate the evolution operator $\mathbb{U}(t)$, which reads
\begin{eqnarray} \label{s18a}
\mathbb{U}(t)&=&\mathbb{U}(t')\mathbb{U}(\mathcal{N}T)=\left[
                \begin{array}{cc}
                 A_n(t')+iB_n(t') & C_n(t')-iD_n(t') \\[1ex]
                 -C_n(t')-iD_n(t') & A_n(t')-iB_n(t') \\
                \end{array}
              \right]\cdot\left[
                \begin{array}{cc}
                 A_N(T)+iB_N(T) & C_N(T)-iD_N(T) \\[1ex]
                 -C_N(T)-iD_N(T) & A_N(T)-iB_N(T) \\
                \end{array}
              \right]^{\mathcal{N}} \cr\cr\cr
              &=&\left[
                \begin{array}{cc}
                 A(t)+iB(t) & C(t)-iD(t) \\[1ex]
                 -C(t)-iD(t) & A(t)-iB(t) \\
                \end{array}
              \right].
\end{eqnarray}
The coefficients $A(t)$, $B(t)$, $C(t)$, and $D(t)$ are given by
\begin{eqnarray} \label{s18b}
A(t)&=& A_n(t')\cos\mathcal{N}\Theta-\left[\mathcal{A}_n(t')\cdot\mathcal{A}^\texttt{T}_N(T)\right]\frac{\sin\mathcal{N}\Theta}{\sin\Theta}, \cr\cr
B(t)&=& B_n(t')\cos\mathcal{N}\Theta+\left[\mathcal{B}_n(t')\cdot\mathrm{diag}\{-1,-1,1\}\cdot\mathcal{A}^\texttt{T}_N(T)\right] \frac{\sin\mathcal{N}\Theta}{\sin\Theta}, \cr\cr
C(t)&=& C_n(t')\cos\mathcal{N}\Theta+\left[\mathcal{C}_n(t')\cdot\mathrm{diag}\{-1,1,-1\}\cdot\mathcal{A}^\texttt{T}_N(T)\right] \frac{\sin\mathcal{N}\Theta}{\sin\Theta}, \cr\cr
D(t)&=& D_n(t')\cos\mathcal{N}\Theta+\left[\mathcal{D}_n(t')\cdot\mathrm{diag}\{1,-1,-1\}\cdot\mathcal{A}^\texttt{T}_N(T)\right] \frac{\sin\mathcal{N}\Theta}{\sin\Theta},
\end{eqnarray}
where $\Theta=\arccos A_N(T)$ and the superscript $\texttt{T}$ denotes the transposition.
It is worth mentioning that both the effective Hamiltonian in Eq.~(\ref{s5}) and the evolution operator in Eq.~(\ref{s18a}) {are exact, since we do not make any approximations.}

\subsection{Example: Two-step sequence}

As an example, let us calculate the expression (\ref{s19}) for the two-step sequence, where the Hamiltonian is
\begin{eqnarray}
H(t)=\left \{
\begin{array}{ll}
    H_{1}=\Delta_{1}|2\rangle\langle 2|+\big[\epsilon_{1}\exp({i\theta_{1}})|1\rangle\langle 2|+\mathrm{H.c.}\big], ~~~~&~~~~~ t\in[nT,\tau_1+nT) \nonumber\\[1.7ex]
    H_{2}=\Delta_{2}|2\rangle\langle 2|+\big[\epsilon_{2}\exp({i\theta_{2}})|1\rangle\langle 2|+\mathrm{H.c.}\big], &~~~~~ t\in[\tau_1+nT,(n+1)T). \\
\end{array}
\right.
\end{eqnarray}
Here, the period is $T=\tau_1+\tau_2$, and $n=0,1,\dots,\mathcal{N}$. The coefficients $A_2(T)$, $B_2(T)$, $C_2(T)$, and $D_2(T)$ are
\begin{eqnarray}  \label{s14}
A_2(T)&=&\cos E_1\tau_1\cos E_2\tau_2-\frac{\Delta_1\Delta_2+4\epsilon_1\epsilon_2\cos(\theta_1-\theta_2)}{4E_1E_2}\sin E_1\tau_1\sin E_2\tau_2, \nonumber\\[0.3ex]
B_2(T)&=&\frac{\Delta_1}{2E_1}\sin E_1\tau_1\cos E_2\tau_2+\left[\frac{\Delta_2}{2E_2}\cos E_1\tau_1 +\frac{\epsilon_1\epsilon_2\sin(\theta_1-\theta_2)}{E_1E_2}\sin E_1\tau_1\right]\sin E_2\tau_2,     \nonumber\\[0.3ex]
C_2(T)&=&\frac{\epsilon_1\sin\theta_1}{E_1}\sin E_1\tau_1\cos E_2\tau_2+\left(\frac{\epsilon_2\sin\theta_2}{E_2}\cos E_1\tau_1+\frac{\Delta_2\epsilon_1\cos\theta_1-\Delta_1\epsilon_2\cos\theta_2}{2E_1E_2}\sin E_1\tau_1\right)\sin E_2\tau_2,    \nonumber\\[0.3ex]
D_2(T)&=&\frac{\epsilon_1\cos\theta_1}{E_1}\sin E_1\tau_1\cos E_2\tau_2+\left(\frac{\epsilon_2\cos\theta_2}{E_2}\cos E_1\tau_1+\frac{\Delta_1\epsilon_2\sin\theta_2-\Delta_2\epsilon_1\sin\theta_1}{2E_1E_2}\sin E_1\tau_1\right)\sin E_2\tau_2.
\end{eqnarray}
Thus, the effective Hamiltonian $H_{\mathrm{eff}}$ for the two-step sequence becomes
\begin{eqnarray}
H_{\mathrm{eff}}=\Delta_{\mathrm{eff}}|2\rangle\langle 2|+\big[\epsilon_{\mathrm{eff}}\exp({i\theta_{\mathrm{eff}}})|1\rangle\langle 2|+\mathrm{H.c.}\big],
\end{eqnarray}
where the effective quantities are
$$\Delta_{\mathrm{eff}}=\frac{2B_2(T)}{T\sqrt{1-A_2(T)^2}}\arccos A_2(T),~~~~~\ \epsilon_{\mathrm{eff}}=\frac{\sqrt{C_2(T)^2+D_2(T)^2}}{T\sqrt{1-A_2(T)^2}}\arccos A_2(T),
~~~~~\ \theta_{\mathrm{eff}}=\arctan\frac{C_2(T)}{D_2(T)}.$$

For the two-step sequence with one jump, the coefficients $A_3(T)$, $B_3(T)$, $C_3(T)$, and $D_3(T)$ are
\begin{eqnarray}
A_3(T)&=&\cos E_1\tau_1\cos E_2\tau_2-\frac{\Delta_1\Delta_2+4\epsilon_1\epsilon_2\cos(\theta_1-\theta_2)}{4E_1E_2}\sin E_1\tau_1\sin E_2\tau_2, \cr\cr
B_3(T)&=&\frac{\Delta_1}{2E_1}\sin E_1\tau_1\cos E_2\tau_2+\frac{\Delta_2}{2E_2}\cos E_1\tau_1\sin E_2\tau_2 +\frac{\epsilon_1\epsilon_2\sin(\theta_1-\theta_2)}{E_1E_2}\sin [(2\lambda-1)E_1\tau_1]\sin E_2\tau_2    \cr\cr
&&+\frac{\Delta_2\epsilon_1^2-\Delta_1\epsilon_1\epsilon_2}{E_1^2E_2}\sin [(1-\lambda)E_1\tau_1]\sin \lambda E_1\tau_1\sin E_2\tau_2,  \cr\cr
C_3(T)&=&\frac{\epsilon_1\sin\theta_1}{E_1}\sin E_1\tau_1\cos E_2\tau_2+\frac{\epsilon_2\sin\theta_2}{E_2}\cos [(1-\lambda)E_1\tau_1]\cos \lambda E_1\tau_1 \sin E_2\tau_2 \cr\cr
&&+\frac{\Delta_2\epsilon_1\cos\theta_1-\Delta_1\epsilon_2\cos\theta_2}{2E_1E_2}\sin [(2\lambda-1)E_1\tau_1]\sin E_2\tau_2     \cr\cr
&&+\frac{\Delta_1^2\epsilon_2\sin\theta_2-2\epsilon_1\Delta_1\Delta_2\sin\theta_1+4\epsilon_1^2\epsilon_2\sin(2\theta_1-\theta_2)} {4E_1^2E_2}\sin [(1-\lambda)E_1\tau_1]\sin \lambda E_1\tau_1\sin E_2\tau_2 , \cr\cr
D_3(T)&=&\frac{\epsilon_1\cos\theta_1}{E_1}\sin E_1\tau_1\cos E_2\tau_2+\frac{\epsilon_2\cos\theta_2}{E_2}\cos [(1-\lambda)E_1\tau_1]\cos \lambda E_1\tau_1 \sin E_2\tau_2 \cr\cr
&&+\frac{\Delta_1\epsilon_2\sin\theta_2-\Delta_2\epsilon_1\sin\theta_1}{2E_1E_2}\sin [(2\lambda-1)E_1\tau_1]\sin E_2\tau_2 \cr\cr
&&+\frac{\Delta_1^2\epsilon_2\cos\theta_2-2\epsilon_1\Delta_1\Delta_2\cos\theta_1+4\epsilon_1^2\epsilon_2\cos(2\theta_1-\theta_2)} {4E_1^2E_2}\sin [(1-\lambda)E_1\tau_1]\sin \lambda E_1\tau_1\sin E_2\tau_2.
\end{eqnarray}
Thus, the effective Hamiltonian $H_{\mathrm{eff}}$ for the two-step sequence with one jump becomes
\begin{eqnarray}
H_{\mathrm{eff}}(\lambda)=\Delta_{\mathrm{eff}}|2\rangle\langle 2|+\big[\epsilon_{\mathrm{eff}}\exp({i\theta_{\mathrm{eff}}})|1\rangle\langle 2|+\mathrm{H.c.}\big],
\end{eqnarray}
where the effective quantities are
$$\Delta_{\mathrm{eff}}=\frac{2B_3(T)}{T\sqrt{1-A_3(T)^2}}\arccos A_3(T),~~~~~\ \epsilon_{\mathrm{eff}}=\frac{\sqrt{C_3(T)^2+D_3(T)^2}}{T\sqrt{1-A_3(T)^2}}\arccos A_3(T),
~~~~~\ \theta_{\mathrm{eff}}=\arctan\frac{C_3(T)}{D_3(T)}.$$

\section{Some other well-known phenomena in the periodic $N$-step driven system }  \label{s3}

In this section, apart from the beat phenomenon mentioned in the main text, we demonstrate that some other well-known phenomena, such as {coherent destruction of tunneling, population complete transition, periodic evolution, stepwise evolution, and population swapping}, can also be recovered in the periodic $N$-step driven (PNSD) system.

Note that we only employ the effective Hamiltonian to study this issue, and thus investigate the dynamical behaviors at the time $t=NT$, where $N$ is an integer. In addition, whether the transition probability can reach unity or not is determined by the quantity $P_l=\epsilon_{\mathrm{eff}}^2/\omega_{\mathrm{eff}}^2$.
The physical parameters should satisfy different conditions for different phenomena, as explained in the following.

\begin{enumerate}
\item {Coherent destruction of tunneling}: $\epsilon_{\mathrm{eff}}=0$, namely, $C_N(T)=D_N(T)=0$.
\item {Complete population transition}: $\Delta_{\mathrm{eff}}=0$, namely, $B_N(T)=0$.
\item {Periodic evolution}: $\cos \mathcal{N}_1\Theta=1$, namely, $\mathcal{N}_1\Theta=\pi$, where $\mathcal{N}_1$ is an integer.
\item {Stepwise evolution}: $\cos \mathcal{N}'_1\Theta=0$, namely, $\mathcal{N}'_1\Theta=\pi/2$, where $\mathcal{N}'_1$ is the steps.
\item {Population swapping}: $\cos \Theta=0$, namely, $\Theta=\pi/2$. Actually, this is the specific stepwise evolution when $\mathcal{N}'_1=1$.
\end{enumerate}

In Figs.~\ref{sfig:03a}--\ref{sfig:03c}, we plot the dynamical evolutions to demonstrate different phenomena by the two-step sequence. Furthermore, the results {again demonstrate} that the actual dynamics of the PNSD system can be well described by only employing $P_{12}^m(t)$ with a few main frequencies, where the expressions can be found in Table \ref{t1}.

\begin{figure}[htbp]
\centering
\includegraphics[scale=0.75]{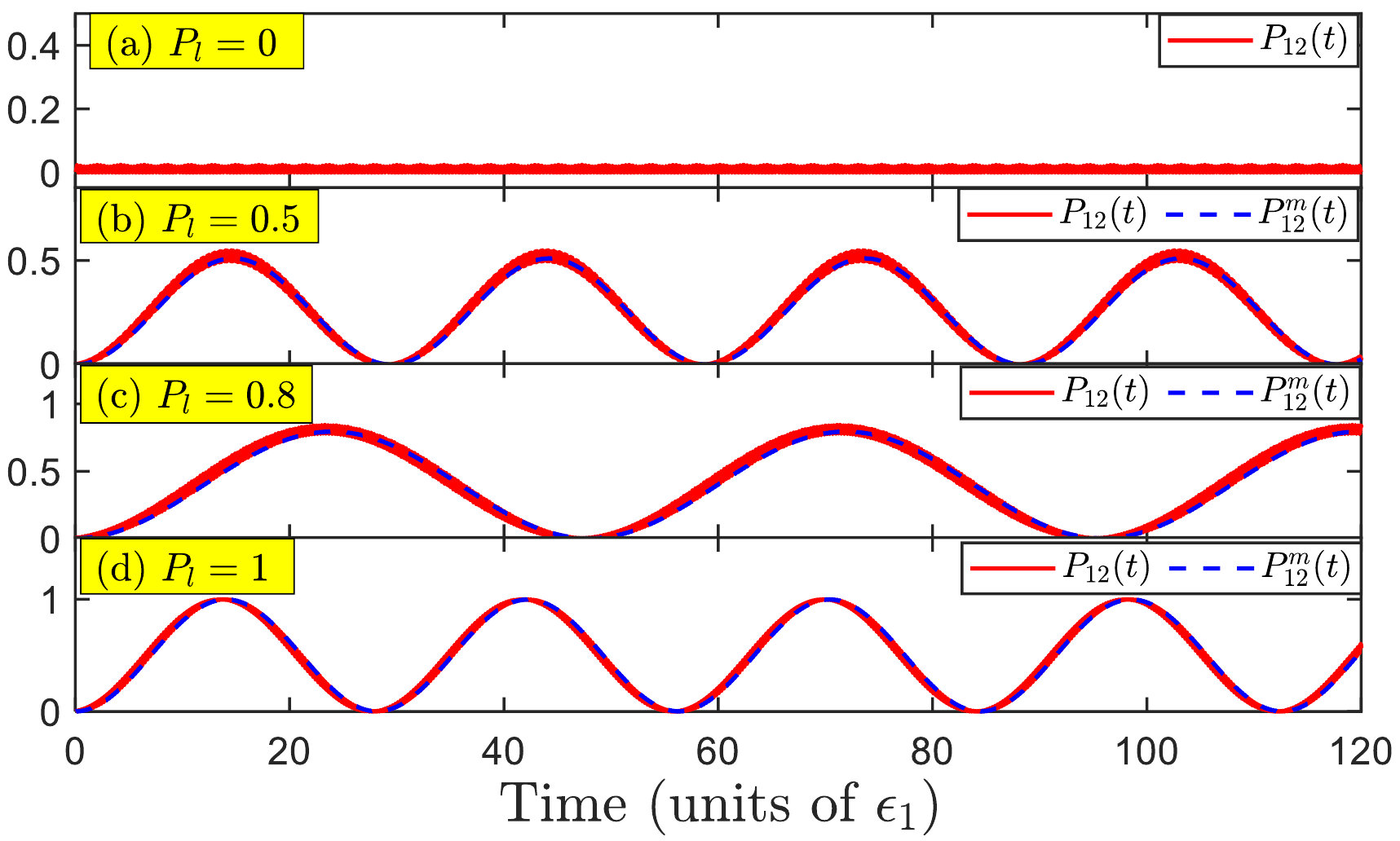}
\caption{ (a) Example of {coherent destruction of tunneling} {for} periodic $N$-step driven (PNSD) systems. The parameters used here are $\{\theta_n\}=\{0,\pi\}$, $\{\vartheta_n\}=\{0.05,0.05\}$, $\{\Delta_n/\epsilon_1\}=\{0,0\}$, and $\epsilon_1=\epsilon_2$. (b-d) Examples of the complete (incomplete) population transition in the PNSD system. The parameters are found in Table \ref{t1}. The red solid curves represent the actual dynamics and the blue dashed curves correspond to the predictions given by $P_{12}^m(t)$. The appreciable overlap between the red solid and the blue dashed curves verifies that the actual dynamics of the PNSD system can be well described by only employing $P_{12}^m(t)$ with a few main frequencies.}  \label{sfig:03a}
\end{figure}

\begin{figure}[htbp]
\centering
\includegraphics[scale=0.76]{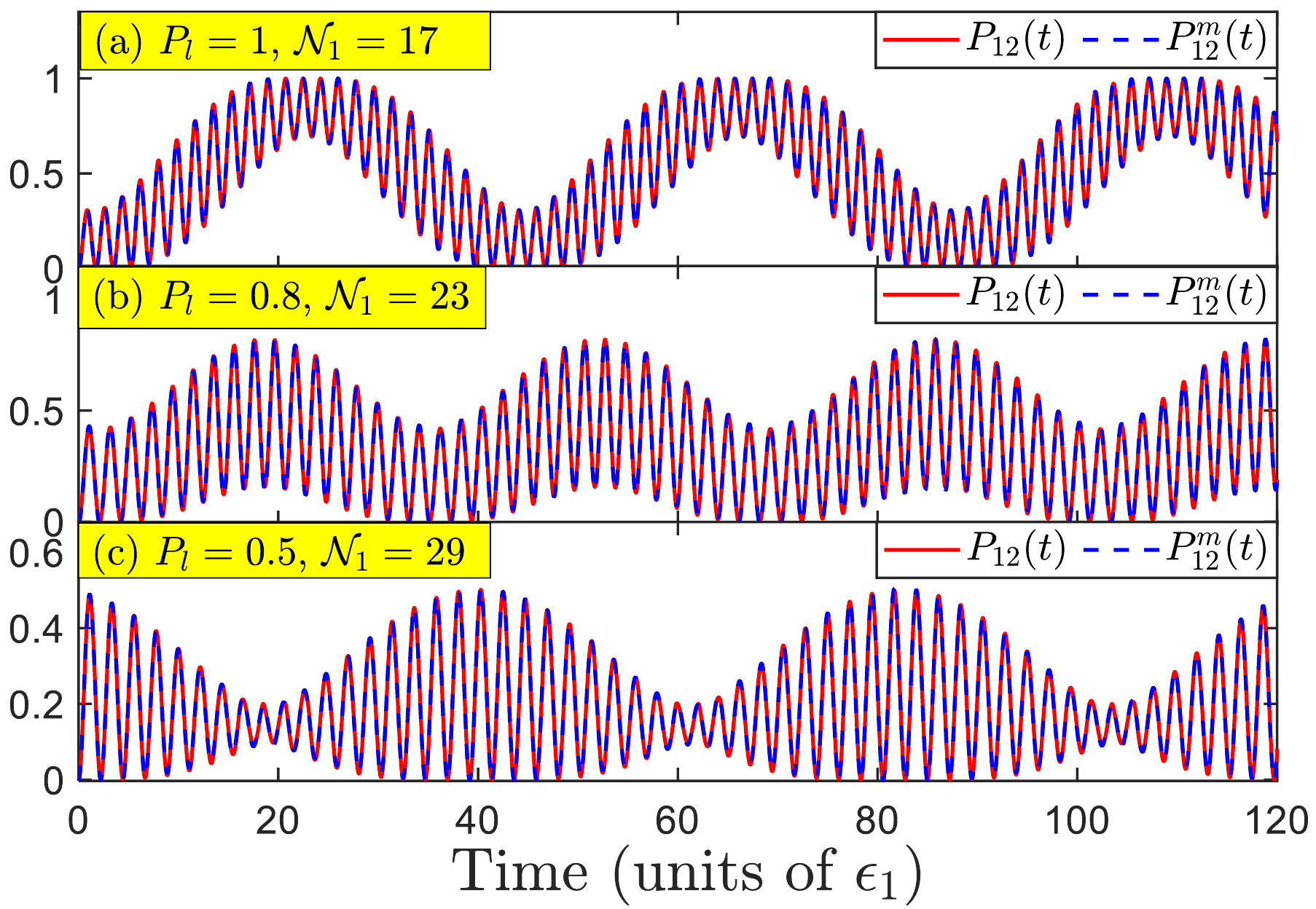}
\caption{ (a) Examples of periodic evolution in the PNSD system. The parameters used here are found in Table \ref{t1}. The appreciable overlap between the red solid and the blue dashed curves verifies that the actual dynamics of the PNSD system can be well described by only employing $P_{12}^m(t)$ with a few main frequencies.}  \label{sfig:03b}
\end{figure}

\begin{figure}[htbp]
\centering
\includegraphics[scale=0.76]{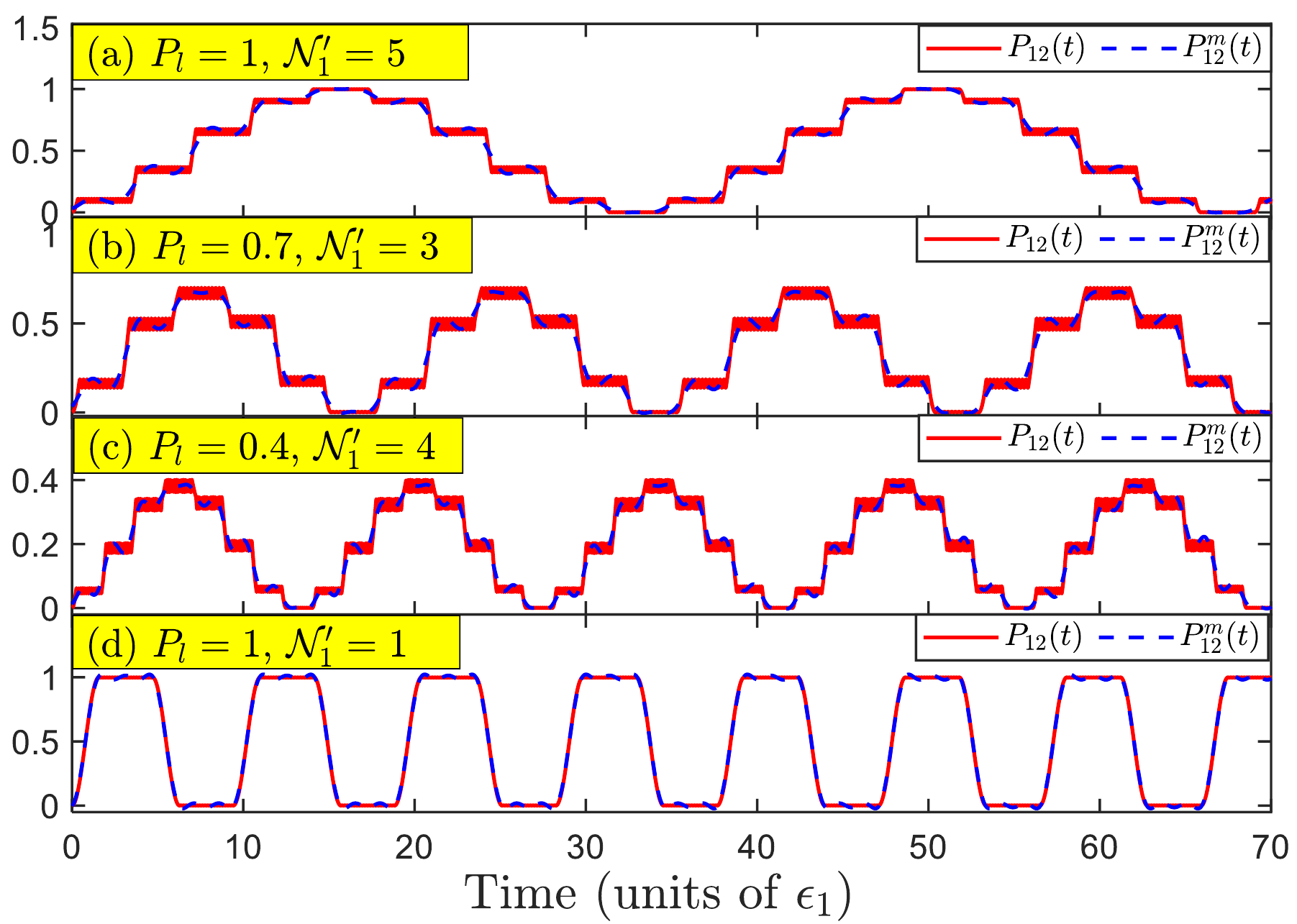}
\caption{ Examples of the step evolution in the PNSD system. The parameters used here are found in Table \ref{t1}. The appreciable overlap between the red solid and the blue dashed curves verifies that the actual dynamics of the PNSD system can be well described by only employing $P_{12}^m(t)$ with a few main frequencies.}  \label{sfig:03c}
\end{figure}

\section{Fourier transform of the time-dependent transition probability $P_{12}(t)$}  \label{ie}

In this section, we consider a \emph{two-step} sequence as an example to demonstrate the Fourier transform of the time-dependent transition probability $P_{12}(t)$. Note that it is easy to generalize to the case of $N>2$. With the evolution operator $\mathbb{U}(t)$ in Eq.~(\ref{s18a}), the time-dependent transition probability $P_{12}(t)$ from state $|1\rangle$ to $|2\rangle$ reads
\begin{eqnarray}
P_{12}(t)=C(t)^2+D(t)^2.
\end{eqnarray}
According to Eq.~(\ref{s18b}), the expressions of $C(t)$ and $D(t)$ can be written as ($k=0,1,2,\dots$)
\begin{eqnarray}
C(t)=\left \{
\begin{array}{ll}
    R^{1c}_{k}\cos E_1t+R^{1s}_{k}\sin E_1t, & t\in[kT,\tau_1+kT), \\[1.7ex]
    Q^{1c}_{k}\cos E_2(t-\tau_1)+Q^{1s}_{k}\sin E_2(t-\tau_1), ~~~~~~~& t\in[\tau_1+kT,(k+1)T), \\
\end{array}
\right.  \cr\cr\cr
D(t)=\left \{
\begin{array}{ll}
    R^{2c}_{k}\cos E_1t+R^{2s}_{k}\sin E_1t, & t\in[kT,\tau_1+kT), \\[1.7ex]
    Q^{2c}_{k}\cos E_2(t-\tau_1)+Q^{2s}_{k}\sin E_2(t-\tau_1), ~~~~~~~& t\in[\tau_1+kT,(k+1)T), \\
\end{array}
\right.
\end{eqnarray}
where the time-independent coefficients are
\begin{eqnarray}
R^{1c}_{k}&=&C_2(T)\frac{\sin k\Theta}{\sin\Theta},  ~~~~~~~~R^{1s}_{k}=\Big(-B_2(T)\frac{\sin k\Theta}{\sin\Theta},\cos k\Theta,D_2(T)\frac{\sin k\Theta}{\sin\Theta}\Big)\cdot\mathcal{E}_1, \nonumber\\[1.2ex]
R^{2c}_{k}&=&D_2(T)\frac{\sin k\Theta}{\sin\Theta},  ~~~~~~~~R^{2s}_{k}=\Big(\cos k\Theta,B_2(T)\frac{\sin k\Theta}{\sin\Theta},-C_2(T)\frac{\sin k\Theta}{\sin\Theta}\Big)\cdot\mathcal{E}_1, \nonumber\\[1.2ex]
Q^{1c}_{k}&=&C_1(\tau_1)\cos k\Theta+\Big[\mathcal{C}_{1}(\tau_1)\cdot\mathrm{diag}\{-1,1,-1\}\cdot\mathcal{A}_2^\texttt{T}(T)\Big]\frac{\sin k\Theta}{\sin\Theta}, \cr\cr
Q^{1s}_{k}&=&\Big[\mathcal{C}_{1}(\tau_1)\cdot\mathcal{E}_2\Big]\cos k\Theta+\Bigg\{ D_2(T)\Big[\mathcal{B}_{1}(\tau_1)\cdot\mathcal{E}_2\Big]-C_2(T)\Big[\mathcal{A}_{1}(\tau_1)\cdot\mathcal{E}_2\Big] -B_2(T)\Big[\mathcal{D}_{1}(\tau_1)\cdot\mathcal{E}_2\Big]\Bigg\}\frac{\sin k\Theta}{\sin\Theta},  \cr\cr
Q^{2c}_{k}&=&D_1(\tau_1)\cos k\Theta+\Big[\mathcal{D}_{1}(\tau_1)\cdot\mathrm{diag}\{1,-1,-1\}\cdot\mathcal{A}_2^\texttt{T}(T)\Big]\frac{\sin k\Theta}{\sin\Theta}, \cr\cr
Q^{2s}_{k}&=&\Big[\mathcal{D}_{1}(\tau_1)\cdot\mathcal{E}_2\Big]\cos k\Theta-\Bigg\{ D_2(T)\Big[\mathcal{A}_{1}(\tau_1)\cdot\mathcal{E}_2\Big]+C_2(T)\Big[\mathcal{B}_{1}(\tau_1)\cdot\mathcal{E}_2\Big] -B_2(T)\Big[\mathcal{C}_{1}(\tau_1)\cdot\mathcal{E}_2\Big]\Bigg\}\frac{\sin k\Theta}{\sin\Theta},  \nonumber
\end{eqnarray}

As a result, the exact solution of the time-dependent transition probability $P_{12}(t)$ from $|1\rangle$ to $|2\rangle$ becomes
\begin{eqnarray} \label{s37b}
P_{12}(t)=C(t)^2+D(t)^2=\left \{
\begin{array}{ll}
    R_k^0+R_k^c\cos 2E_1t+R_k^s\sin 2E_1t, & t\in[kT,\tau_1+kT), \\\cr
    Q_k^0+Q_k^c\cos 2E_2(t-\tau_1)+Q_k^s\sin 2E_2(t-\tau_1), ~~~~~~~& t\in[\tau_1+kT,(k+1)T), \\
\end{array}
\right.
\end{eqnarray}
where  the coefficients are
\begin{eqnarray}
R_k^0&=&\frac{(R^{1c}_{k})^2+(R^{1s}_{k})^2+(R^{2c}_{k})^2+(R^{2s}_{k})^2}{2}, ~~~~~R_k^c=\frac{(R^{1c}_{k})^2-(R^{1s}_{k})^2+(R^{2c}_{k})^2-(R^{2s}_{k})^2}{2},
~~~~~R_k^s=R^{1c}_{k}R^{1s}_{k}+R^{2c}_{k}R^{2s}_{k}, \cr\cr
Q_k^0&=&\frac{(Q^{1c}_{k})^2+(Q^{1s}_{k})^2+(Q^{2c}_{k})^2+(Q^{2s}_{k})^2}{2},
~~~~~Q_k^c=\frac{(Q^{1c}_{k})^2-(Q^{1s}_{k})^2+(Q^{2c}_{k})^2-(Q^{2s}_{k})^2}{2},
~~~~~Q_k^s=Q^{1c}_{k}Q^{1s}_{k}+Q^{2c}_{k}Q^{2s}_{k}.    \nonumber
\end{eqnarray}

Although $P_{12}(t)$ in Eqs.~(\ref{s37b}) is a quasiperiodic function rather than a periodic function, the frequency spectrum is still discrete and the fundamental frequency is $\omega_T=2\pi/T$.
Moreover, the frequency spectrum is also associated with the effective Rabi frequency $\omega_\mathrm{eff}=\sqrt{\epsilon_\mathrm{eff}^2+{\Delta_\mathrm{eff}^2}/{4}}$.
By the Fourier transform, the expression of $P_{12}(t)$ can be expanded in terms of the cosine functions, yielding
\begin{eqnarray}    \label{s29}
P_{12}(t)&&=\bar{b}+\sum_{l=-\infty}^{\infty}\Big\{b_{l}\cos\big[(2\omega_{\mathrm{eff}}+l\omega_{T})t-\varphi_l\big] +b_{l}^{\prime}\cos\left[(l+1)\omega_{T}t-\varphi_l^{\prime}\right]\Big\},
\end{eqnarray}
where $b_{l}=\sqrt{(c_{l})^2+(d_{l})^2}$, $b_{l}^{\prime}=\sqrt{(c_{l}^{\prime})^2+(d_{l}^{\prime})^2}$, $\varphi_l=\arctan[c_{l}/d_{l}]$, $\varphi_l^{\prime}=\arctan[c_{l}^{\prime}/d_{l}^{\prime}]$, and
\begin{eqnarray}
\bar{b}&=&\lim_{K\rightarrow\infty}\frac{1}{KT}\int_0^{KT}P_{12}(t)dt,  \cr\cr
c_{l}&=&\lim_{K\rightarrow\infty}\frac{2}{KT}\int_0^{KT}P_{12}(t)\sin(2\omega_{\mathrm{eff}}
+l\omega_{T})tdt, \cr\cr
c_{l}^{\prime}&=&\lim_{K\rightarrow\infty}\frac{2}{KT}\int_0^{KT}P_{12}(t)\sin(l+1)\omega_{T}tdt,  \cr\cr
d_{l}&=&\lim_{K\rightarrow\infty}\frac{2}{KT}\int_0^{KT}P_{12}(t)\cos(2\omega_{\mathrm{eff}}
+l\omega_{T})tdt,  \cr\cr
d_{l}^{\prime}&=&\lim_{K\rightarrow\infty}\frac{2}{KT}\int_0^{KT}P_{12}(t)\cos(l+1)\omega_{T}tdt.  \nonumber
\end{eqnarray}

In order to derive those coefficients, one needs to calculate the following integrals:
\begin{eqnarray}
I_{l}(t_2,t_1,R^c,R^s,\omega_{\mathrm{eff}},\tau_1)&\equiv&\int_{t_1}^{t_2}\Big[R^c\cos(2\omega_{\mathrm{eff}}+l\omega_T)(t-\tau_1) +R^s\sin(2\omega_{\mathrm{eff}}+l\omega_T)(t-\tau_1)\Big]dt \cr\cr
&=&\frac{R^c}{2\omega_{\mathrm{eff}}+l\omega_T} \Big[\sin(2\omega_{\mathrm{eff}}+l\omega_T)(t_2-\tau_1)-\sin(2\omega_{\mathrm{eff}}+l\omega_T)(t_1-\tau_1)\Big] \cr\cr &&-\frac{R^s}{2\omega_{\mathrm{eff}}+l\omega_T}\Big[\cos(2\omega_{\mathrm{eff}}+l\omega_T)(t_2-\tau_1) -\cos(2\omega_{\mathrm{eff}}+l\omega_T)(t_1-\tau_1)\Big],  \cr\cr
I_{l}^c(t_2,t_1,R^c,R^s,\omega_{\mathrm{eff}},E_2,\tau_1) &\equiv&\int_{t_1}^{t_2}\Big[R^c\cos2E_2(t-\tau_1)+R^s\sin2E_2(t-\tau_1)\Big]\cos(2\omega_{\mathrm{eff}}+l\omega_T)tdt  \cr\cr
&=&R^c\frac{\sin[(2\omega_{\mathrm{eff}}+l\omega_T+2E_2)t_2-2E_2\tau_1] -\sin[(2\omega_{\mathrm{eff}}+l\omega_T+2E_2)t_1-2E_2\tau_1]}{2(2\omega_{\mathrm{eff}}+l\omega_T+2E_2)} \cr\cr &&+R^c\frac{\sin[(2\omega_{\mathrm{eff}}+l\omega_T-2E_2)t_2+2E_2\tau_1] -\sin[(2\omega_{\mathrm{eff}}+l\omega_T-2E_2)t_1+2E_2\tau_1]}{2(2\omega_{\mathrm{eff}}+l\omega_T-2E_2)} \cr\cr
&&+R^s\frac{\cos[(2\omega_{\mathrm{eff}}+l\omega_T+2E_2)t_1-2E_2\tau_1] -\cos[(2\omega_{\mathrm{eff}}+l\omega_T+2E_2)t_2-2E_2\tau_1]}{2(2\omega_{\mathrm{eff}}+l\omega_T+2E_2)} \cr\cr &&-R^s\frac{\cos[(2\omega_{\mathrm{eff}}+l\omega_T-2E_2)t_1+2E_2\tau_1] -\cos[(2\omega_{\mathrm{eff}}+l\omega_T-2E_2)t_2+2E_2\tau_1]}{2(2\omega_{\mathrm{eff}}+l\omega_T-2E_2)}, \cr\cr
I_{l}^s(t_2,t_1,R^c,R^s,\omega_{\mathrm{eff}},E_2,\tau_1) &\equiv&\int_{t_1}^{t_2}\Big[R^c\cos2E_2(t-\tau_1)+R^s\sin2E_2(t-\tau_1)\Big]\sin(2\omega_{\mathrm{eff}}+l\omega_T)tdt  \cr\cr
&=&R^c\frac{\cos[(2\omega_{\mathrm{eff}}+l\omega_T+2E_2)t_1-2E_2\tau_1] -\cos[(2\omega_{\mathrm{eff}}+l\omega_T+2E_2)t_2-2E_2\tau_1]}{2(2\omega_{\mathrm{eff}}+l\omega_T+2E_2)}  \cr\cr &&+R^c\frac{\cos[(2\omega_{\mathrm{eff}}+l\omega_T-2E_2)t_1+2E_2\tau_1] -\cos[(2\omega_{\mathrm{eff}}+l\omega_T-2E_2)t_2+2E_2\tau_1]}{2(2\omega_{\mathrm{eff}}+l\omega_T-2E_2)}  \cr\cr
&&+R^s\frac{\sin[(2\omega_{\mathrm{eff}}+l\omega_T-2E_2)t_2+2E_2\tau_1] -\sin[(2\omega_{\mathrm{eff}}+l\omega_T-2E_2)t_1+2E_2\tau_1]}{2(2\omega_{\mathrm{eff}}+l\omega_T-2E_2)} \cr\cr &&-R^s\frac{\sin[(2\omega_{\mathrm{eff}}+l\omega_T+2E_2)t_2-2E_2\tau_1] -\sin[(2\omega_{\mathrm{eff}}+l\omega_T+2E_2)t_1-2E_2\tau_1]}{2(2\omega_{\mathrm{eff}}+l\omega_T+2E_2)}. \nonumber
\end{eqnarray}
Then, the coefficients are
\begin{eqnarray}   \label{b4}
\bar{b}&=&\lim_{K\rightarrow\infty}\frac{1}{KT}\int_0^{KT}[C(t)^2+D(t)^2]dt,  \cr
&=&\lim_{K\rightarrow\infty}\frac{1}{KT}\sum_{k=0}^{K-1}\int_{kT}^{kT+\tau_1}[C(t)^2+D(t)^2]dt+ \frac{1}{KT}\sum_{k=0}^{K-1}\int_{kT+\tau_1}^{(k+1)T}[C(t)^2+D(t)^2]dt \cr
&=&\lim_{K\rightarrow\infty}\frac{1}{KT}\sum_{k=0}^{K-1}\int_{kT}^{kT+\tau_1}(R_k^0+R_k^c\cos2E_1t+R_k^s\sin2E_1t)dt  \cr &&+\lim_{K\rightarrow\infty}\frac{1}{KT}\sum_{k=0}^{K-1}\int_{kT+\tau_1}^{(k+1)T} [Q_k^0+Q_k^c\cos2E_2(t-\tau_1)+Q_k^s\sin2E_2(t-\tau_1)]dt \cr
&=&\lim_{K\rightarrow\infty}\frac{1}{KT}\sum_{k=0}^{K-1}\Big[R_k^0\tau_1+Q_k^0\tau_2+I_{0}(kT+\tau_1,kT,R_k^c,R_k^s,E_1,0) +I_{0}(kT+T,kT+\tau_1,Q_k^c,Q_k^s,E_2,\tau_1)\Big],  \cr
c_{l}&=&\lim_{K\rightarrow\infty} \frac{2}{KT}\sum_{k=0}^{K-1}\Big[I_{l}(kT+\tau_1,kT,0,R_k^0,\omega_{\mathrm{eff}},0) +I_{l}(kT+T,kT+\tau_1,0,Q_k^0,\omega_{\mathrm{eff}},0)    \cr
&&+I_{l}^s(kT+\tau_1,kT,R_k^c,R_k^s,\omega_{\mathrm{eff}},E_1,kT) +I_{l}^s(kT+T,kT+\tau_1,Q_k^c,Q_k^s,\omega_{\mathrm{eff}},E_2,\tau_1+kT) \Big],    \cr
c_{l}^{\prime}&=&\lim_{K\rightarrow\infty}\frac{2}{KT}\sum_{k=0}^{K-1}\Big[I_{l+1} (kT+\tau_1,kT,0,R_k^0,0,0)+I_{l+1}(kT+T,kT+\tau_1,0,Q_k^0,0,0)    \cr
&&+I_{l+1}^s(kT+\tau_1,kT,R_k^c,R_k^s,0,E_1,kT) +I_{l+1}^s(kT+T,kT+\tau_1,Q_k^c,Q_k^s,0,E_2,\tau_1+kT) \Big],    \cr
d_{l}&=&\lim_{K\rightarrow\infty}\frac{2}{KT}\sum_{k=0}^{K-1}\Big[I_{l}(kT+\tau_1,kT,R_k^0,0,\omega_{\mathrm{eff}},0) +I_{l}(kT+T,kT+\tau_1,Q_k^0,0,\omega_{\mathrm{eff}},0)    \cr
&&+I_{l}^c(kT+\tau_1,kT,R_k^c,R_k^s,\omega_{\mathrm{eff}},E_1,kT) +I_{l}^c(kT+T,kT+\tau_1,Q_k^c,Q_k^s,\omega_{\mathrm{eff}},E_2,\tau_1+kT) \Big],    \cr
d_{l}^{\prime}&=&\lim_{K\rightarrow\infty}\frac{2}{KT}\sum_{k=0}^{K-1}\Big[I_{l+1}(kT+\tau_1,kT,R_k^0,0,0,0)+I_{l+1}(kT+T,kT+\tau_1,Q_k^0,0,0,0)    \cr
&&+I_{l+1}^c(kT+\tau_1,kT,R_k^c,R_k^s,0,E_1,kT) +I_{l+1}^c(kT+T,kT+\tau_1,Q_k^c,Q_k^s,0,E_2,\tau_1+kT) \Big].
\end{eqnarray}

Note that all the integrals $I_l(\cdot)$, $I_l^c(\cdot)$, and $I_l^s(\cdot)$ are superpositions of trigonometric functions.  By making use of the summation formulas of trigonometric functions
\begin{eqnarray}
\sum_{k=0}^{K}\cos kx=\frac{\sin\frac{x}{2}+\sin(\frac{1}{2}+K)x}{2\sin\frac{x}{2}},~~~~~~~~~~~~~~~~~~~~
\sum_{k=0}^{K}\sin kx=\frac{\cos\frac{x}{2}-\cos(\frac{1}{2}+K)x}{2\sin\frac{x}{2}},~~~
\end{eqnarray}
one {can obtain} the limits {for} concrete physical parameters.

As an example, consider $\theta_1=\theta_2=0$, $\Delta_1=0$, ${|\Delta_2|}\gg{|\epsilon_2|}$,
and $E_1\tau_1=E_2\tau_2=\pi/2$. Substituting those parameters into the expression~(\ref{s37b}),  we have
\begin{eqnarray}
P_{12}(t)=\left \{
\begin{array}{ll}
    \displaystyle\frac{1}{2}-\frac{1}{2}\cos2k\Theta\cos 2E_1t, & t\in[kT,\tau_1+kT), \\[1.7ex]
    \sin^2(k+1)\Theta\sin^2E_2(t-\tau_1)+\cos^2k\Theta\cos^2E_2(t-\tau_1), ~~~~~~~& t\in[\tau_1+kT,(k+1)T), \\
\end{array}
\right.
\end{eqnarray}
where $\Theta=\arccos\epsilon_2/E_2$. Then, according to Eq.~(\ref{b4}), the coefficients in the Fourier transform become
\begin{eqnarray}
\bar{b}&=&\lim_{K\rightarrow\infty}\frac{1}{KT}\int_0^{KT}P_{12}^b(t)dt=\frac{1}{2},  \nonumber\\[0.8ex]
c_{0}&=&\lim_{K\rightarrow\infty}\frac{2}{KT}\int_0^{KT}P_{12}^b(t)\sin(2\omega_{\mathrm{eff}}
)tdt\simeq\frac{1}{4}\frac{\Theta[1+\cos\frac{2\tau_1}{T}\Theta]}{({\pi T}/{2\tau_1})^2-\Theta^2}, \nonumber\\[0.8ex]
d_{0}&=&\lim_{K\rightarrow\infty}\frac{2}{KT}\int_0^{KT}P_{12}^b(t)\cos(2\omega_{\mathrm{eff}})tdt \simeq-\frac{1}{4}\frac{\Theta\sin\frac{2\tau_1}{T}\Theta}{({\pi T}/{2\tau_1})^2-\Theta^2}, \nonumber\\[0.8ex]
c_{-1}&=&\lim_{K\rightarrow\infty}\frac{2}{KT}\int_0^{KT}P_{12}^b(t)\sin(2\omega_{\mathrm{eff}}
-\omega_{T})tdt\simeq\frac{1}{4}\frac{(\pi-\Theta)[1+\cos\frac{2\tau_1}{T}(\pi-\Theta)]}{({\pi T}/{2\tau_1})^2-(\pi-\Theta)^2}, \nonumber\\[0.8ex]
d_{-1}&=&\lim_{K\rightarrow\infty}\frac{2}{KT}\int_0^{KT}P_{12}^b(t)\cos(2\omega_{\mathrm{eff}}-\omega_{T})tdt \simeq-\frac{1}{4}\frac{(\pi-\Theta)\sin\frac{2\tau_1}{T}(\pi-\Theta)}{({\pi T}/{2\tau_1})^2-(\pi-\Theta)^2}.   \nonumber
\end{eqnarray}
Hence, $b_{0}=\sqrt{(c_{0})^2+(d_{0})^2}\simeq1/4$, $b_{-1}=\sqrt{(c_{-1})^2+(d_{-1})^2}\simeq1/4$. Therefore, $P_{12}(t)$ can be approximated by
\begin{eqnarray}   \label{s34a}
P_{12}(t)\simeq P_{12}^m(t)=\frac{1}{2}-\frac{1}{4}\cos(2\omega_{\mathrm{eff}}t)-\frac{1}{4}\cos(2\omega_{\mathrm{eff}}^{-}t),
\end{eqnarray}
where $\omega_{\mathrm{eff}}^{-}=\omega_T/2-\omega_{\mathrm{eff}}$. One finds that the transition probability shows the beat phenomenon in this parameter regime. Indeed, this is simply what Eq.~(\ref{s37a}) gives when we choose $N=2$.

\section{Beat phenomenon in periodic $N$-step driven {systems}}   \label{id}

In the main text, we demonstrate that the beat phenomenon can emerge in the periodic \emph{$N$-step} driven system.
Without loss of generality, here we set $\theta_n=0$ and study the situation where the beat phenomenon is most obvious. That is, there exists a complete transition in the periodic $N$-step driven system ($\Delta_{\mathrm{eff}}=0$), and the transition probability can be approximately given by
\begin{eqnarray}  \label{s37a}
P_{12}^{{b}}(t)=\frac{1}{2}\left[\sin^2\varpi_1(t-t_p) +\sin^2\varpi_1'(t-t_p)\right],
\end{eqnarray}
where the frequencies $\varpi_1$ and $\varpi_1'$ are closely related to the physical quantities of the periodic $N$-step driven system, and $t_p$ is {a} time-shifting factor.

First of all, we need to estimate the sum of all dynamical phases. If all Hamiltonians are in the largely detuned regime ($\left|{\Delta_n}/{\epsilon_n}\right|\gg1$, $n=1,2,\dots,N$), the following relations hold:
\begin{eqnarray}
E_n=\sqrt{\epsilon_n^2+{\Delta_n^2}/{4}}&\simeq&{\Delta_n}/{2},\nonumber\\[0.8ex]
{\epsilon_n}/{E_n}&\simeq&0.
\end{eqnarray}
Substituting into the equation $B_N(T)=0$, we have $\sum_{n=1}^{N}E_n\tau_n\simeq k\pi$, $k=1,2,\dots$.
As a result, to implement a complete transition between levels, the accumulation of all dynamical phases approximately equal to an integer multiple of $\pi$ if all Hamiltonians are in the largely detuned regime.
When $N$ is an even number, the choice of physical quantities is quite obvious. That is, we can set each dynamical phase to be $E_n\tau_n={\pi}/{2}$. Then we have
\begin{eqnarray}
\sum_{n=1}^{N}E_n\tau_n={N\pi}/{2},
\end{eqnarray}
which satisfies the complete transition condition.
At the same time, in order for the beating to emerge, one should guarantee that at least one Hamiltonian is in the resonant regime and the remaining Hamiltonians are in the largely detuned regime.
Then, the expressions of the frequencies $\varpi_1$ and $\varpi_1'$ are empirically given by
\begin{eqnarray}  \label{s37}
\varpi_1=\frac{1}{2}\left \{
\begin{array}{ll}
    {(n_1-1)\omega^{-}_{\mathrm{eff}}+(n_1+1)\omega_{\mathrm{eff}}}, ~& \mathrm{odd}~n_{1}, \\[1.7ex]
    {n_1\omega^{-}_{\mathrm{eff}}+(n_1-2)\omega_{\mathrm{eff}}}, ~& \mathrm{even}~n_{1}, \\
\end{array}
\right.
\end{eqnarray}
and
\begin{eqnarray}    \label{s38}
\varpi_1'=\frac{1}{2}\left \{
\begin{array}{ll}
    {(n_1+1)\omega^{-}_{\mathrm{eff}}+(n_1-1)\omega_{\mathrm{eff}}}, ~& \mathrm{odd}~n_{1}, \\[1.7ex]
    {n_1\omega^{-}_{\mathrm{eff}}+(n_1+2)\omega_{\mathrm{eff}}},  ~& \mathrm{even}~n_{1}, \\
\end{array}
\right.
\end{eqnarray}
where $n_{1}$ represents the number of Hamiltonians that are in the resonant regime.

Indeed, the beat phenomenon always exists in the periodic $N$-step driven system. Taking the two-step sequence as an example, let us estimate the values of the frequencies $\varpi_1$ and $\varpi_1'$. Assume that the Hamiltonian $H_1$ is in the resonant regime and the Hamiltonian $H_2$ is in the largely detuned regime. We have $\Delta_1=0$ and $|\Delta_2/\epsilon_2|\gg1$. Substituting those parameters into expression (\ref{s14}), we find that the value of $\varpi_1$ can  be approximated by
\begin{eqnarray}
\varpi_1=\omega_{\mathrm{eff}}\approx\frac{\arccos A_2(T)}{T}=\frac{1}{T}{\arccos \left[\frac{-\epsilon_2\cos(\theta_1-\theta_2)}{E_2}\right]}= \frac{1}{T}{\arccos \left[\frac{-\epsilon_2\cos(\theta_1-\theta_2)}{\sqrt{\epsilon_2^2+\Delta_2^2/4}}\right]}\approx\frac{\pi}{2T},
\end{eqnarray}
since $\epsilon_2/\sqrt{\epsilon_2^2+\Delta_2^2/4}\approx0$.
Then, the value of $\varpi'_1$ can be approximated by
\begin{eqnarray}
\varpi'_1=\omega_{\mathrm{eff}}^{-}=\frac{\omega_T}{2}-\omega_{\mathrm{eff}}\approx\frac{\pi}{T}-\frac{\pi}{2T}=\frac{\pi}{2T}.
\end{eqnarray}
As a result, once $\cos(\theta_1-\theta_2)\neq0$, the frequencies $\varpi_1$ and $\varpi_1'$ are different but similar. Moreover, we have derived the expression of the transition probability in Eq.~(\ref{s34a}), which demonstrates that the amplitudes are approximately equal to each other. Thus, the beating always exists in this system. Using \textcolor[rgb]{0.00,0.00,1.00}{a} similar derivation, one can obtain the same results when $N>2$.

We plot in Fig.~\ref{sfig:08}(a) the average error $\varepsilon^{\mathrm{m}}(t_s)$ as a function of $\Delta_6$ for different $n_1$.
These results demonstrate that Eq.~(\ref{s37a}) with the frequencies $\varpi_1$ and $\varpi_1'$ given by Eqs. (\ref{s37}) and (\ref{s38}) can be used to describe well the actual dynamics of the periodic $N$-step driven system, because the maximum value of $\varepsilon^{\mathrm{m}}(t_s)$ is not larger than 0.066. As examples, we plot in Figs.~\ref{sfig:08}(b) and \ref{sfig:08}(c) the actual and predicted dynamical evolution of the system when $n_1=1$ and $n_1=2$, respectively.

\begin{figure}[htbp]
\centering
\includegraphics[scale=0.62]{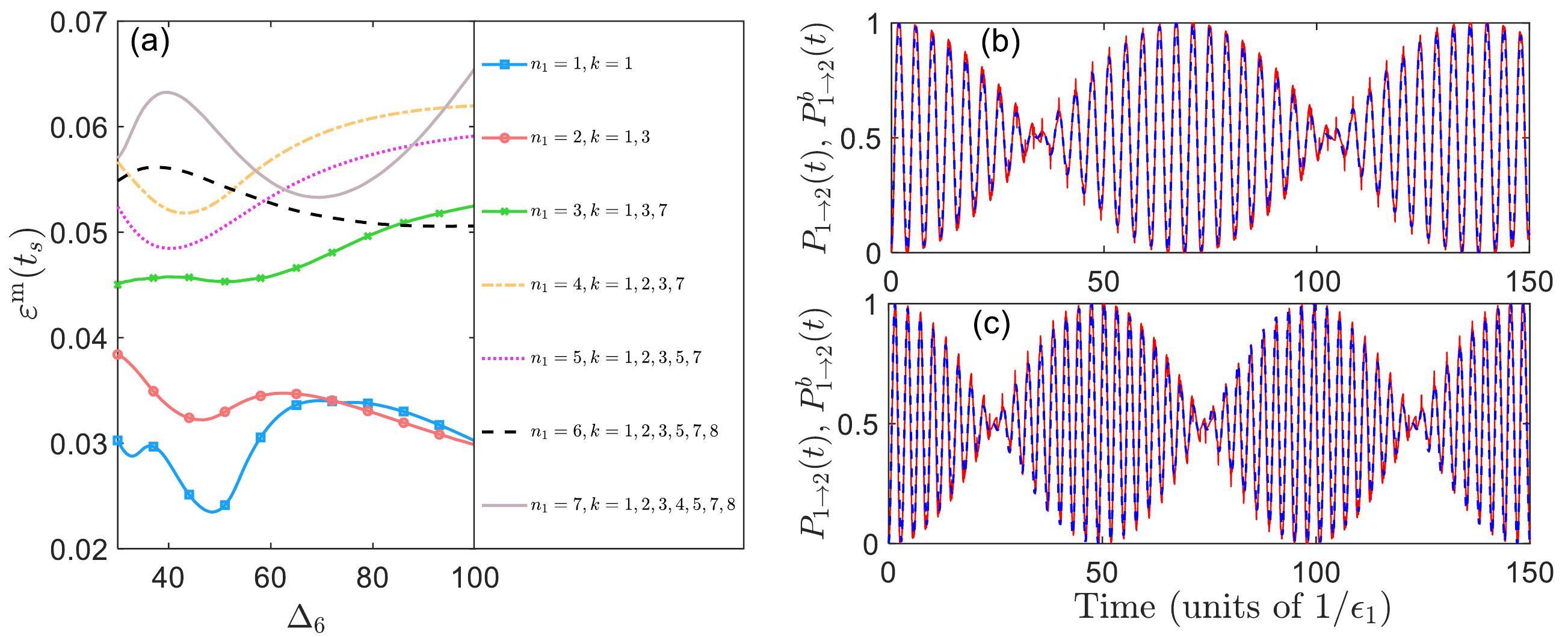}
\caption{ (a) $\varepsilon^{\mathrm{m}}(t_s)$ vs $\Delta_6$ with different $n_1$ in a periodic $N$-step driven system, where $N=8$ and we properly select a set of physical quantities (in units of $\epsilon_1$): $\epsilon_2=1.8$, $\epsilon_3=1.5$, $\epsilon_4=1.1$, $\epsilon_5=1.2$, $\epsilon_6=1.6$, $\epsilon_7=1.9$,  $\epsilon_8=1.3$, $\Delta_1=0$,  $\Delta_2=67$, $\Delta_3=60$,  $\Delta_4=46$, $\Delta_5=83$, $\Delta_6=36$, $\Delta_7=40$, and $\Delta_8=91$.
In the legend, each $k$ indicates that the $k$th Hamiltonian is in the resonant regime.
(b, c) Time-dependent transition probabilities $P_{12}(t)$ with (b) $n_1=1$ and (c) $n_1=2$ in a periodic $N$-step driven system. The red solid curves represent the actual dynamics and the blue dashed curves correspond to the predictions given by Eq.~(\ref{s37a}). Those results show that when $N$ is even, the actual dynamics can be well described by the beating formula (\ref{s37a}) with the frequencies (\ref{s37}) and (\ref{s38}). }  \label{sfig:08}
\end{figure}

When $N$ is an odd number, the choice of physical quantities is slightly different from the even $N$ case. That is, the dynamical phase caused by one Hamiltonian $H_{k_0}$ should satisfy $E_{k_0}\tau_{k_0}=\pi$, while the remaining Hamiltonians still satisfy $E_n\tau_n={\pi}/{2}$ ($n\neq k_0$).
Thus we can consider two cases according to the Hamiltonian $H_{k_0}$:

(i) The Hamiltonian $H_{k_0}$ is in the largely detuned regime. One easily finds that
the evolution operator of the Hamiltonian $H_{k_0}$ is the identity operator; thus it does not affect the system dynamics. On the other hand, due to the largely detuned regime, the transition probability caused by this Hamiltonian is small enough. Thus, we can ignore this Hamiltonian and regard the $N$-step driven system as an $(N-1)$-step driven system. As a result, the expressions of {the} coefficients $\varpi_1$ and $\varpi_1'$ in this case {are} the same as in the situation where $N$ is an odd number, which is governed by Eqs. (\ref{s37}) and (\ref{s38}).

(ii) The Hamiltonian $H_{k_0}$ is in the resonant regime. In this case, the dynamical phase caused by the Hamiltonian $H_{k_0}$ is twice the one caused by the Hamiltonian in the situation where $N$ is an odd number. Therefore, when the number of Hamiltonians that are in the resonant regime is assumed to be $n_1$, we should substitute $(n_1+1)$ into Eqs. (\ref{s37}) and (\ref{s38}) to achieve the expressions of the frequencies $\varpi_1$ and $\varpi_1'$. Consequently, in this case the expressions can be written as
\begin{eqnarray}  \label{s39}
\varpi_1=\frac{1}{2}\left\{
\begin{array}{ll}
    (n_1+1)\omega^{-}_{\mathrm{eff}}+(n_1-1)\omega_{\mathrm{eff}}, ~~~~~& n_1~\textrm{is odd} \\[1.7ex]
    n_1\omega^{-}_{\mathrm{eff}}+(n_1+2)\omega_{\mathrm{eff}}, ~~~~~& n_1~\textrm{is even}, \\
\end{array}
\right.
\end{eqnarray}
and
\begin{eqnarray}   \label{s40}
\varpi_1'=\frac{1}{2}\left\{
\begin{array}{ll}
    (n_1+1)\omega^{-}_{\mathrm{eff}}+(n_1+3)\omega_{\mathrm{eff}}, ~& n_1~\textrm{is odd} \\[1.7ex]
    (n_1+2)\omega^{-}_{\mathrm{eff}}+n_1\omega_{\mathrm{eff}},  ~& n_1~\textrm{is even}. \\
\end{array}
\right.
\end{eqnarray}
We also plot in Fig.~\ref{sfig:09}(a) the average error $\varepsilon^{\mathrm{m}}(t_s)$ as a function of $\Delta_7$ with different $n_1$.
It demonstrates that Eq.~(\ref{s37a}) with the frequencies $\varpi_1$ and $\varpi_1'$ given by Eqs. (\ref{s39}) and (\ref{s40}) can be used to  describe well the actual dynamics of the periodic $N$-step driven system, since the maximum value of $\varepsilon^{\mathrm{m}}(t_s)$ is not larger than 0.082. As examples, we plot in Figs.~\ref{sfig:09}(b) and \ref{sfig:09}(c) the actual and predicted dynamical evolution of the system when $n_1=2$ and $n_1=3$, respectively.

\begin{figure}[htbp]
\centering
\includegraphics[scale=0.62]{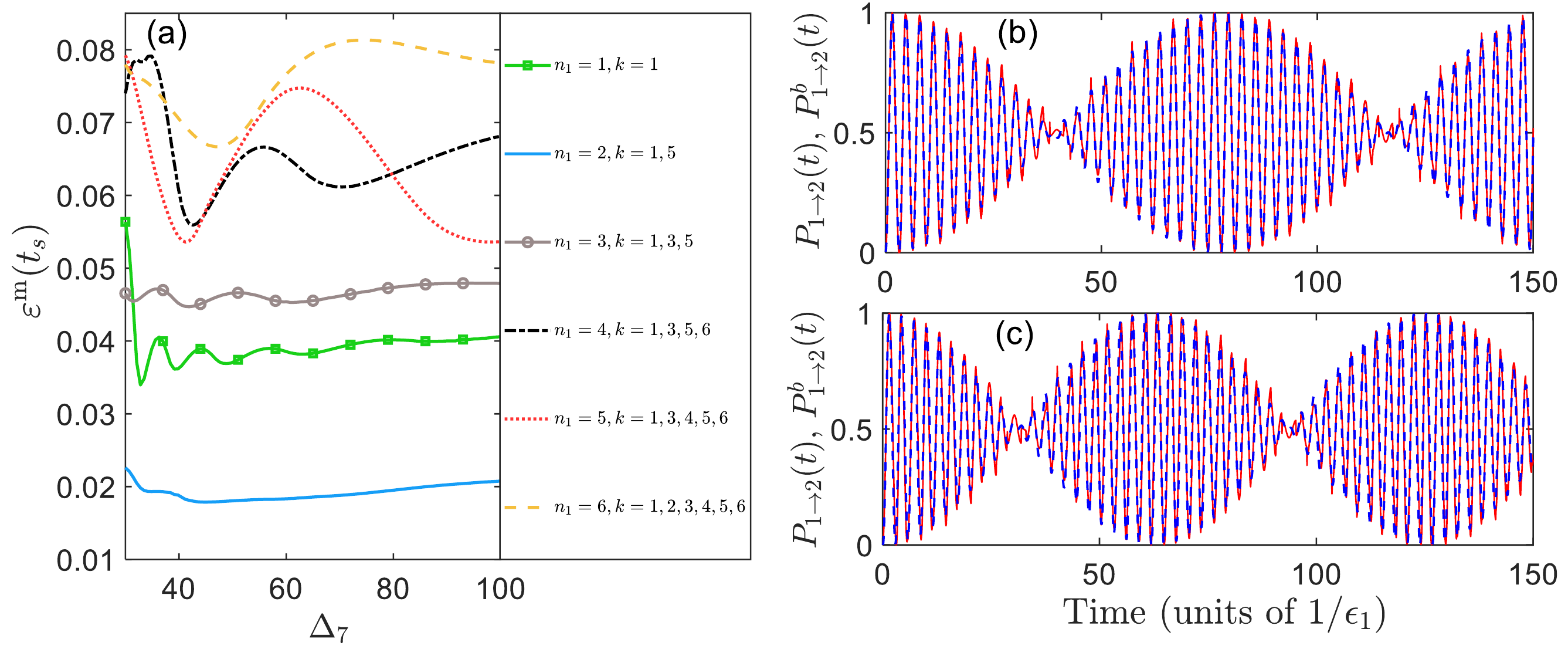}
\caption{ (a) $\varepsilon^{\mathrm{m}}(t_s)$ vs $\Delta_6$ with different $n_1$ in a periodic $N$-step driven system, where  $N=7$ and we properly select a set of physical quantities (in units of $\epsilon_1$): $\epsilon_2=1.8$, $\epsilon_3=1.5$, $\epsilon_4=1.1$, $\epsilon_5=1.2$, $\epsilon_6=1.6$, $\epsilon_7=1.9$, $\epsilon_8=1.3$, $E_{1}\tau_{1}=\pi$, $\Delta_1=0$,  $\Delta_2=55$, $\Delta_3=62$,  $\Delta_4=47$, $\Delta_5=88$, $\Delta_6=35$, and $\Delta_7=65$. In the legend, each $k$ indicates that the $k$th Hamiltonian is in the resonant regime.
(b, c) Time-dependent transition probabilities $P_{12}(t)$ with (b) $n_1=2$ and (c) $n_1=3$ in the periodic $N$-step driven system. The red solid curves represent the actual dynamics and the blue dashed curves correspond to the predictions given by Eq.~(\ref{s37a}). Those results show that when $N$ is odd, the actual dynamics can be well described by the beating formula (\ref{s37a}) with the frequencies (\ref{s39}) and (\ref{s40}). }  \label{sfig:09}
\end{figure}

\end{widetext}

\end{appendix}

\bibliography{references}

\end{document}